\documentstyle[emulateapj,epsf]{article}
\newcommand{\nab}{\mbox{\boldmath $\nabla$}} 
\newcommand{\xib}{\mbox{\boldmath $\xi$}} 
\newcommand{\xibc}{\mbox{\boldmath $\xi^*$}} 
\newcommand{\xibt}{\mbox{\boldmath $\tilde{\xi}$}}
\newcommand{\xibct}{\mbox{\boldmath $\tilde{\xi}^*$}}  
\newcommand{\omegab}{\mbox{\boldmath $\Omega$}} 
\newcommand{\timesb}{\mbox{\boldmath $\times$}} 
\newcommand{\be}{\begin{equation}} 
\newcommand{\ee}{\end{equation}} 
\newcommand{\ben}{\begin{eqnarray}} 
\newcommand{\een}{\end{eqnarray}} 
\newcommand{\pl}{\partial} 
\newcommand{\gau}{\left(} 
\newcommand{\dr}{\right)}

\newcommand{\veck}{\mbox{\boldmath $\hat{k}$}} 
 
\received{1997 November 17} 
\revised{} 
\accepted{1998 July 24} 
\cpright{}{}
\journalid{}{} 
\articleid{}{} 
\paperid{} 
 
\cpright{}{} 
\ccc{} 
 
\slugcomment{The Astrophysical Journal, 509:000--000, 1998 December 20} 
 
\lefthead{ACCRETION DISKS AND BENDING WAVES} 
\righthead{TERQUEM} 
 
\begin{document} 

\title{The response of accretion disks to bending waves: angular
momentum transport and resonances}
 
\author{Caroline E.J.M.L.J. Terquem\footnote{On leave from Laboratoire
d'Astrophysique, Observatoire de Grenoble, Universit\'e Joseph
Fourier/CNRS, 38041 Grenoble Cedex 9, France}}
 
\affil{UCO/Lick Observatory, University of California, Santa~Cruz, CA
95064; Isaac Newton Institute for Mathematical Sciences, University of
Cambridge, 20 Clarkson Road, Cambridge,~CB3~0EH, UK; ct@ucolick.org }

\begin{abstract} 
 
We investigate the linear tidal perturbation of a viscous Keplerian
disk by a companion star orbiting in a plane inclined to the disk.  We
consider $m=1$ perturbations with odd symmetry with respect to the
$z=0$ midplane.  The response frequency may be either finite or
vanishing.  These long--wavelength perturbations produce a
well--defined warp.  

Since the response of a viscous disk is not in phase with the
perturbing potential, a tidal torque is exerted on the disk. When the
perturber rotates outside the disk, this torque results in a decrease
of the disk angular momentum, and thus in an increase of its accretion
rate. 

We show that this tidal torque is comparable to the horizontal viscous
stress acting on the background flow when the perturbed velocities in
the disk are on the order of the sound speed.  If these velocities
remain subsonic, the tidal torque can exceed the horizontal viscous
stress only if the viscous parameter $\alpha_{\rm v}$ which couples to
the vertical shear is larger than the parameter $\alpha_{\rm h}$
coupled to the horizontal shear. In protostellar disks, bending waves,
which are predominantly excited in the outer regions, are found to
propagate and transport a significant fraction of the negative angular
momentum they carry deep into the disk inner parts.  

If the waves are reflected at the center, resonances occur when the
frequency of the tidal waves is equal to that of some free normal
global bending mode of the disk.  If such resonances exist, tidal
interactions may then be important even when the binary separation is
large.  Out of resonance, the torque associated with the secular
perturbation, which is proportional to $\alpha_{\rm v}$, is generally
much larger than that associated with the finite frequency
perturbations.  As long as the waves are damped before they reach the
center, the torque associated with the finite frequency perturbations
does not depend on the viscosity, in agreement with theoretical
expectation. These calculations are relevant to disks around young
stars and maybe also to disks in X--ray binary systems.
 
\end{abstract} 
 
\subjectheadings{accretion, accretion disks --- binaries: general ---
hydrodynamics --- \\ stars: pre--main sequence --- waves}
 
\section{Introduction} 

\subsection{Tidal Interactions in Pre--main--Sequence Binary Systems: 
Coplanar versus Noncoplanar Systems}
\label{sec:intro1}

Thanks to recent optical images taken with the {\it Hubble Space
Telescope}, accretion disks around low--mass pre--main sequence stars
known as T~Tauri are no longer objects of theoretical speculations
(McCaughrean~\& O'Dell~1996).  As expected, they appear to be common
around young stars (Stauffer et al.~1994). Furthermore, most T~Tauri
stars are observed to be in multiple systems (see Mathieu~1994 and
references therein). Since the distribution of the separation of
pre--main sequence binaries has a peak around 30 astronomical units,
which is smaller than the extent of the disks in such systems (Edwards
et al.~1987), it is expected that many circumstellar disks will be
subject to tidal effects due to the influence of binary companions.
 
The theory of tidal interaction has already successfully satisfied
many observational tests. For example, it has been predicted that
angular momentum exchange between orbital motion and disk rotation
through tidal interaction may truncate the disk (see, e.g.,
Papaloizou~\& Pringle~1977; Paczy\'nski~1977). This has recently
received observational confirmation, both through the study of young
binaries spectra (Dutrey, Guilloteau~\& Simon~1994; Jensen, Mathieu~\&
Fuller~1996) and direct imaging (Roddier et al.~1996).
 
The tidal effect of an orbiting body on a differentially rotating disk
has been well studied in the context of planetary rings (Goldreich~\&
Tremaine~1978), planet formation, and interacting binary stars (see
Lin~\& Papaloizou~1993 and references therein). In these studies, the
disk and orbit are usually taken to be coplanar. However, there are
observational indications that the plane of the disk and that of the
orbit may not necessarily be aligned. The most striking evidence for
such noncoplanarity is given by HST and adaptive optics images of
the pre--main sequence binary HK~Tau (Stapelfeldt~et al.~1998).  The
projected binary separation in this system is 340~AU, the observed
radius of the circumsecondary disk is 105~AU and the disk aspect ratio
at a radius of 50~AU is about 0.08.  The authors suggest that the
primary would be located at an angle of at least $20{\arcdeg}$ above
the disk plane.  In the case of the hierarchical triple system Ty~Cra,
some models suggest that the orbit of the close spectroscopic binary
and that of the wider binary are not in the same plane (see Corporon,
Lagrange~\& Beust~1996; Bibo, The~\& Dawanas~1992). Also, observations
of the pre--main sequence binary T~Tau have revealed that two bipolar
outflows of very different orientations originate from this system
(B\"ohm~\& Solf~1994). Since it is unlikely that they are both ejected
by the same star, each of them is more probably associated with a
different member of the binary.  Furthermore, jets are usually thought
to be ejected perpendicularly to the disk from which they
originate. These observations then suggest that disks and orbit are
not coplanar in this binary system. More generally, observations of
molecular outflows in star forming regions commonly show several jets
of different orientations emanating from unresolved regions of extent
as small as a few hundred astronomical units (Davis, Mundt~\&
Eisl\"offel~1994). 

We note that the spectral energy distribution of a circumstellar disk
could be significantly affected by the lack of coplanarity in binary
or multiple systems (Terquem~\& Bertout~1993, 1996). This is because
reprocessing of radiation from the central star by a disk depends
crucially on the disk geometry.

In such noncoplanar systems, both density and bending waves are
excited in the disk by the perturbing companion. They are respectively
of even and odd symmetry with respect to reflection in the disk
mid--plane. Since much work has already been carried out on density
waves, which are the only waves excited in coplanar systems, we shall
focus here on bending waves. Because their wavelength is larger than
that of density waves, they are expected to propagate and transport
angular momentum further radially into the disk.

\subsection{Bending Waves}
 
Bending waves were first studied by Hunter~\& Toomre~(1969) in the
context of the Galactic disk. They calculated the dispersion relation
associated with these waves taking into account self--gravity but
ignoring pressure and viscosity. Bending waves were further considered
in the context of galactic disks (Toomre~1983; Sparke~1984; Sparke~\&
Casertano~1988) and planetary rings (Shu, Cuzzi~\& Lissauer~1983;
Shu~1984). In these studies, the horizontal motions associated with
the vertical displacement of the disk were neglected. However,
Papaloizou~\& Pringle~(1983) showed that these motions are important
in gaseous Keplerian disks, where they are nearly resonant. These
authors pointed out that when the disk is sufficiently viscous so that
its response to vertical displacements is diffusive rather than
wave--like, these motions lead to the decay timescale of the warp
being $\alpha_{\rm v}^2$ times the viscous timescale $t_{\nu , {\rm
v}}$. Here $\alpha_{\rm v}$ is the Shakura~\& Sunyaev~(1973) parameter
that couples to the vertical shear, and $t_{\nu , {\rm v}}$ is the
disk viscous timescale associated to $\alpha_{\rm v}$.  When
$\alpha_{\rm v}$ is smaller than unity, this timescale is smaller than
$t_{\nu , {\rm v}}$, which is the decay timescale of the warp obtained
by previous authors.

In the context of protostellar disks, Artymowicz~(1994) studied the
orbital evolution of low--mass bodies crossing disks, and
Ostriker~(1994) calculated the angular momentum exchange between an
inviscid disk and a perturber on a near--parabolic noncoplanar orbit
outside the disk (see also Heller~1995 for numerical simulations). The
vertical structure of the disk was not taken into account in these
studies. Papaloizou~\& Lin~(1995) performed a full 3 dimensional
linear analysis of $m=1$ bending waves in accretion disks. Comparing
analytical results with numerical simulations, they showed that a
vertical averaging approximation can be used when the wavelength of
the perturbation is much larger than the disk semi--thickness. Using
the WKB approximation, they calculated the dispersion relation
associated with bending waves taking into account self--gravity,
pressure and viscosity (see also Masset~\& Tagger~1996 for a study of
thick disks). They found that in a non self--gravitating thin
Keplerian disk in which $\alpha_{\rm v}$ is small compared to the disk
aspect ratio, bending waves propagate without dispersion at a speed
which is about half the sound speed.  Using these results,
Papaloizou~\& Terquem~(1995) calculated the $m=1$ bending wave
response of an inviscid disk with the rotation axis misaligned with a
binary companion's orbital rotation axis. This work was extended by
Larwood~et al.~(1996) who performed nonlinear calculations of the disk
response using smoothed particle hydrodynamics (SPH). This numerical
method was further used by Larwood~\& Papaloizou~(1997) who analyzed
the response of a circumbinary disk subject to the tidal forcing of a
binary with a fixed noncoplanar circular orbit, and by Larwood~(1997)
who considered the tidal interaction occurring between a disk and a
perturber following a parabolic trajectory inclined with respect to
the disk.

The present work should be contrasted with earlier studies of purely
viscous disks in which pressure and self--gravity were ignored. The
behavior of nonplanar purely viscous accretion disk subject to
external torques has been considered previously by a number of
authors. Following the analysis of Bardeen~\& Petterson~(1975), these
studies have been developed in connection with the precessing disks of
Her~X--1 and SS~433 for instance (e.g., Petterson~1977; Katz~1980a,
1980b), galactic warps (e.g., Steiman--Cameron~\& Durisen~1988), or
warped disks around AGN and in X--ray binaries
(Pringle~1996). However, the response of a purely viscous disk is very
different from that of a viscous gaseous disk, which is a better
representation of a protostellar disk. In the former case, evolution
occurs only through viscous diffusion, whereas in the latter case
pressure effects manifesting themselves through bending waves can
control the disk evolution. When these waves propagate (i.e.  when
they are not damped), they do so on a timescale comparable to the
sound crossing time. Since this is much shorter than the viscous
diffusion timescale, communication through the disk occurs as a result
of the propagation of these waves.  Even when the waves are damped
before reaching the disk boundary, a diffusion coefficient for warps
that is much larger than that produced by the kinematic viscosity may
occur because of pressure effects (Papaloizou~\& Pringle~1983).
Self--gravity may also strongly modify the response of viscous disks,
because like pressure it allows bending waves to propagate and its
effect is global. For a more detailed discussion of the respective
effects of viscosity, pressure and self--gravity, see Papaloizou,
Terquem~\& Lin~(1998).

\subsection{Torque Exerted on the Disk by a Companion on an Inclined 
Orbit}

There can be no tidal torque exerted on an inviscid disk with
reflective boundaries that contains no corotation resonance
(Goldreich~\& Nicholson~1989).  However, if the disk is viscous or has
a nonreflective boundary, the companion exerts a torque on the disk
that leads to an exchange of angular momentum between the disk
rotation and the orbital motion.  We can distinguish the following
three components of the torque:

1. The component along the line of nodes makes the disk precess at a
rate that can be uniform (see \S~\ref{sec:zero} and Kuijken~1991;
Papaloizou~\& Terquem~1995; Papaloizou et al. 1998).  Although we will
calculate in this paper the precession frequency induced by the tidal
torque, we will not discuss the disk precession in detail since this
has already been done in the papers cited above.

2. The component of the torque along the axis perpendicular to the
line of nodes in the disk plane induces the evolution of the
inclination angle of the disk plane with respect to that of the orbit.
This evolution, which is not necessarily toward coplanarity, occurs on
the disk's viscous timescale if the warp is linear (Papaloizou~\&
Terquem~1995; Papaloizou et al. 1998).  Since it affects the geometry
of the system only on such a long timescale, we will not discuss this
effect here.  We note that both these components of the tidal torque
modify the direction of the disk angular momentum vector.

3. The component of the torque along the disk's spin axis does not
modify the direction but the modulus of the disk's angular momentum
vector, thus changing the disk's accretion rate.  This effect is
important since it is a source of angular momentum transport in the
disk, and it will be the main subject of this paper.

So far, angular momentum exchange between a {\it viscous gaseous} disk
rotation and the orbital motion of a companion on an inclined circular
orbit has not been considered.  It is the goal of this paper to
present an analysis of this process.  In the case of an inviscid disk
considered by Papaloizou~\& Terquem~(1995), such angular momentum
exchange occurs because the waves are assumed to interact nonlinearly
with the background flow before reaching the disk's inner boundary.
Thus, only in-going waves propagate through the disk.  This induces a
lag between the disk response and the perturbation, which enables a
torque to be exerted on the disk.  In a viscous disk, the lag required
for the torque to be non zero is naturally produced by the viscous
shear stress.

\subsection{Plan of the Paper}
 
The plan of the paper is as follows.  In \S~\ref{sec:potential} we
derive an expression for the perturbing potential.  In
\S~\ref{sec:basic} we outline the basic equations.  The model of the
unperturbed disk is described in \S~\ref{sec:equil}.  The disk
response is presented in \S~\ref{sec:response}, where we give the
linearized equations, study the disk response for both zero and finite
perturbing frequencies, establish the boundary conditions and describe
the domain of validity of this analysis.  In \S~\ref{sec:momentum} we
calculate the torque exerted by the perturber on the disk, and relate
it to the disk viscosity by using the conservation of angular
momentum.  We show that this tidal torque is comparable to the
horizontal viscous stress acting on the background flow when the
perturbed velocities in the disk are on the order of the sound speed.
If these velocities remain subsonic, the tidal torque can exceed the
horizontal viscous stress only if the viscous parameter $\alpha_{\rm
v}$ that couples to the vertical shear is larger than the parameter
$\alpha_{\rm h}$ coupled to the horizontal shear.  Numerical results
are presented in \S~\ref{sec:numerical}, both for uniform and
nonuniform viscosity.  They indicate that, for parameters typical of
protostellar disks, bending waves are able to propagate and transport
a significant fraction of the negative angular momentum they carry
deep into the inner parts of the disk.  They also show that, if the
waves can be reflected at the disk's inner boundary, resonances occur
when the frequency of the tidal waves is equal to that of some free
normal global bending mode of the disk.  Out of resonance, if the disk
and the orbital planes are not too close to being perpendicular, the
torque associated with the zero--frequency perturbing term is found to
be much larger than that associated with the finite frequency terms.
As long as the waves are damped before they reach the center, the
torque associated with the finite frequency perturbations does not
depend on the viscosity, in agreement with theoretical expectation
(Goldreich~\& Tremaine~1982).  We summarize and discuss our results in
\S~\ref{sec:discussion}.
 
\section{Perturbing potential} 
\label{sec:potential} 
 
We consider a binary system in which the primary has a mass $M_p$ and 
the secondary has a mass $M_s$. The binary orbit is circular with 
separation $D$. We suppose that the primary is surrounded by a disk of 
radius $R \ll D$ and of negligible mass $M$ so that the orbital plane 
does not precess and the secondary describes a prograde Keplerian 
orbit with angular velocity:
 
\begin{displaymath} 
\omega=\sqrt{ \frac{G \left( M_s+M_p \right)}{D^3}} ,  
\end{displaymath} 
 
\noindent about the primary. In the absence of the secondary star, the 
disk is nearly Keplerian. We adopt a nonrotating Cartesian 
coordinate system $(x,y,z)$ centered on the primary star. The 
$z$--axis is chosen to be normal to the initial disk mid--plane. We 
shall also use the associated cylindrical polar coordinates 
$(r,\varphi,z)$. The orbit of the secondary star is in a plane that 
has an initial inclination angle $\delta$ with respect to the $(x,y)$ 
plane. 
 
We adopt an orientation of coordinates and an origin of time such that
the line of nodes coincides with, and the secondary is on, the
$x$--axis at time $t=0$. We denote the position vector of the
secondary star by $\bf{D}$ with $D \equiv |\bf{D}|$.
 
The total gravitational potential $\Psi$ due to the binary at a point
with position vector $\bf{r}$ is given by:
 
\be 
\Psi = - \frac{GM_p}{\left| \bf{r} \right|}  
- \frac{GM_s}{\left| \bf{r}-\bf{D} \right|}  
+ \frac{GM_s \bf{r} \cdot \bf{D}}{D^3} , 
\label{Psit} 
\ee 
   
\noindent where $G$ is the gravitational constant. The first dominant 
term is due to the primary, while the remaining perturbing terms are 
due to the secondary. The last indirect term accounts for the 
acceleration of the origin of the coordinate system. 
 
We are interested in disk warps that are excited by terms in the
potential that are odd in $z$ and have azimuthal mode number $m=1$
when a Fourier analysis in $\varphi$ is carried out. The terms in the
expansion of the potential that are of the required form are given
by:

\begin{eqnarray} 
\Psi'& = & \frac{\sin \varphi}{2 \pi} \int_0^{2 \pi} \left[
\Psi \left( r, \varphi', z, \omega t \right) \right.
\nonumber \\
& - & \left. \Psi \left( r, \varphi', -z, \omega t \right) \right]
\sin \varphi' d \varphi' .
\end{eqnarray}
 
We consider a thin disk such that $z \ll r$. Since $r \ll D$, we then
retain only the lowest order terms in $r/D$ and $z/D$ in the above
integral, which becomes:
 
\begin{displaymath} 
\Psi'=- \frac{3}{4} \frac{GM_s}{D^3} rz \left[  
\left( 1- \cos \delta \right) \sin \delta  
\sin \left( \varphi + 2 \omega t \right) \right.
\end{displaymath}
\be
\left. - \left( 1+ \cos \delta \right) \sin \delta  
\sin \left( \varphi - 2 \omega t \right) 
+ \sin 2 \delta  \sin \varphi \right] . 
\label{tPsi'} 
\ee 
 
Because the principle of linear superposition can be used, the general
problem may be reduced to calculating the response to a complex
potential of the form:
 
\be 
\Psi'= ifrz e^{i \left(\varphi-\Omega_p t \right)}.  
\label{f} 
\ee 
 
\noindent Here the pattern frequency $\Omega_p$ of the perturber is 
one of $0,$ $2 \omega$ or $-2 \omega$, and $f$ is an appropriate real 
amplitude. 
 
\section{Basic equations} 
\label{sec:basic} 
 
The response of the disk is determined by the equation of continuity:
 
\be 
\frac{\pl \rho}{\pl t} + \nab \cdot \gau \rho {\bf v} \dr = 0 ,
\label{continuity} 
\ee 
 
\noindent and the equation of motion:
 
\be \frac{\partial \bf{v}}{\partial t} + \left( \bf{v} \cdot \nab
\right) {\bf{v}} = \frac{1}{\rho} \left( - \nab P + {\bf{F}}_{\nu}
\right) - \nab \Psi ,
\label{motion} 
\ee 
 
\noindent where $P$ is the pressure, $\rho$ is the mass density and
${\bf v}$ is the flow velocity. We allow for the possible presence of
a viscous force ${\bf{F}}_{\nu}$ but, following Papaloizou~\&
Lin~(1995), we shall assume that it does not affect the undistorted
axisymmetric disk so that it operates on the perturbed flow only.
This approximation is justified because the disk viscous timescale is
much longer than the characteristic timescales of the processes we are
interested in here (in particular, bending waves propagate with a
velocity comparable to the sound speed, so that the waves crossing
time is much shorter than the viscous timescale).
 
For simplicity, we adopt a polytropic equation of state:
 
\begin{equation} 
P=K \rho^{1+1/n} , 
\label{polyt} 
\end{equation}  
 
\noindent where $K$ is the polytropic constant and $n$ is the
polytropic index. The sound speed is then given by $c_s^2=dP/d\rho.$
 
\section{Equilibrium structure of the disk} 
\label{sec:equil} 
 
We adopt the same model as in Papaloizou~\& Terquem~(1995). Namely we 
suppose that the equilibrium disk is axisymmetric and in a state of 
differential rotation such that, in cylindrical coordinates, the flow 
velocity is given by ${\bf{v}}=\left( 0,r\Omega,0 \right).$ For a 
barotropic equation of state, the angular velocity $\Omega$ is a 
function of $r$ alone. 
 
For a thin disk under the influence of a point mass, integration of
the vertical hydrostatic equilibrium equation gives (Papaloizou~\&
Terquem~1995):
 
\begin{equation} 
H(r)=\left( \frac{ \left[ 2K \left( 1+n \right) \right]^n }{C_n}
\right)^{1/ \left( 2n+1 \right)} \left( \Sigma \Omega_K^{-2n}
\right)^{1/ \left( 2n+1 \right)} , 
\label{Hr} 
\end{equation}
 
\begin{displaymath}
\rho(r,z)= \left[ 2K \left( 1+n \right) C_n^2 \right]^{-n/ \left( 2n+1
\right)} 
\end{displaymath}
\begin{equation}
\times \left( \Sigma \Omega_K \right)^{2n/ \left( 2n+1 \right)}
\left( 1- \frac{z^2}{H^2} \right)^n .  
\end{equation}
 
\noindent Here $\Omega_K$ is the Keplerian angular velocity given by 
$\Omega_K^2=GM_p/r^3$, $H$ is the disk semi--thickness, $\Sigma$ is the 
surface mass density defined by:
 
\begin{displaymath} 
\Sigma(r)=\int_{-H(r)}^{H(r)} \rho \left( r,z \right) dz , 
\end{displaymath} 
 
\noindent and  
 
\begin{displaymath} 
C_n=\frac{\Gamma \left( \frac{1}{2} \right) \Gamma \left( n+1 
\right)}{\Gamma \left( \frac{1}{2}+n+1 \right)} , 
\end{displaymath} 
 
\noindent where $\Gamma$ denotes the gamma function. 
 
It then follows from the radial hydrostatic equilibrium equation that 
the angular velocity is given by: 
 
\begin{displaymath}
\Omega^2=\Omega_K^2 + \frac{1}{r} \left\{ \frac{ \left[ K \left( 1+n
\right) \right]^{2n}} {2 C_n^2} \right\}^{1/ \left( 2n+1 \right)}
\end{displaymath}
\begin{equation}
\times \frac{\partial}{\partial r} \left[ \left( \Sigma \Omega_K
\right)^{2/ \left( 2n+1 \right) } \right] .
\label{angul} 
\end{equation}
 
In principle, the surface density profile should be determined 
self--consistently as a result of viscous diffusion (Lynden--Bell~\& 
Pringle~1974). However, since we ignore the effect of viscosity on the 
unperturbed disk, we are free to specify arbitrary density profiles 
for the equilibrium disk model. For $\Sigma$ we then take a function 
that is constant in the main body of the disk with a taper to make it 
vanish at the outer boundary: 
 
\begin{equation} 
\Sigma(r) = \Sigma_0 \frac{\sigma(r)}{\left[ 1+ \sigma(r)^q
\right]^{1/q}} , 
\end{equation}
 
\noindent which approximates $\Sigma(r)=\Sigma_0 \; \rm{min} \left[ 1, 
\sigma(r) \right]$ with 
 
\begin{displaymath}  
\sigma(r) = \left\{ 1 - \left[ \frac{r- \left( R - \Delta \right) 
}{\Delta} \right]^p \right\}^N .   
\end{displaymath} 
 
\noindent The parameters $\Delta$, which represents the width of the
taper interior to the outer boundary, $N$, $q$ and $p$ are constrained
such that the square of the epicyclic frequency, which is defined by:
 
\begin{displaymath} 
\kappa^2 = \frac{2 \Omega}{r} \frac{d \left( r^2 \Omega \right)}{dr} , 
\end{displaymath} 
 
\noindent is positive everywhere in the disk. For the numerical 
calculations we take $N=\left( 2n+1 \right) /2$, $\Delta/R=0.1$, $q=4$ 
and $p=1$. 
 
The polytropic constant $K$ is determined by specifying a maximum
value of the relative semi--thickness, $H/r,$ in the disk, and the
fiducial surface mass density $\Sigma_0$ is fixed by specifying the
total mass $M$ of the disk.
 
\begin{figure}
\plotone{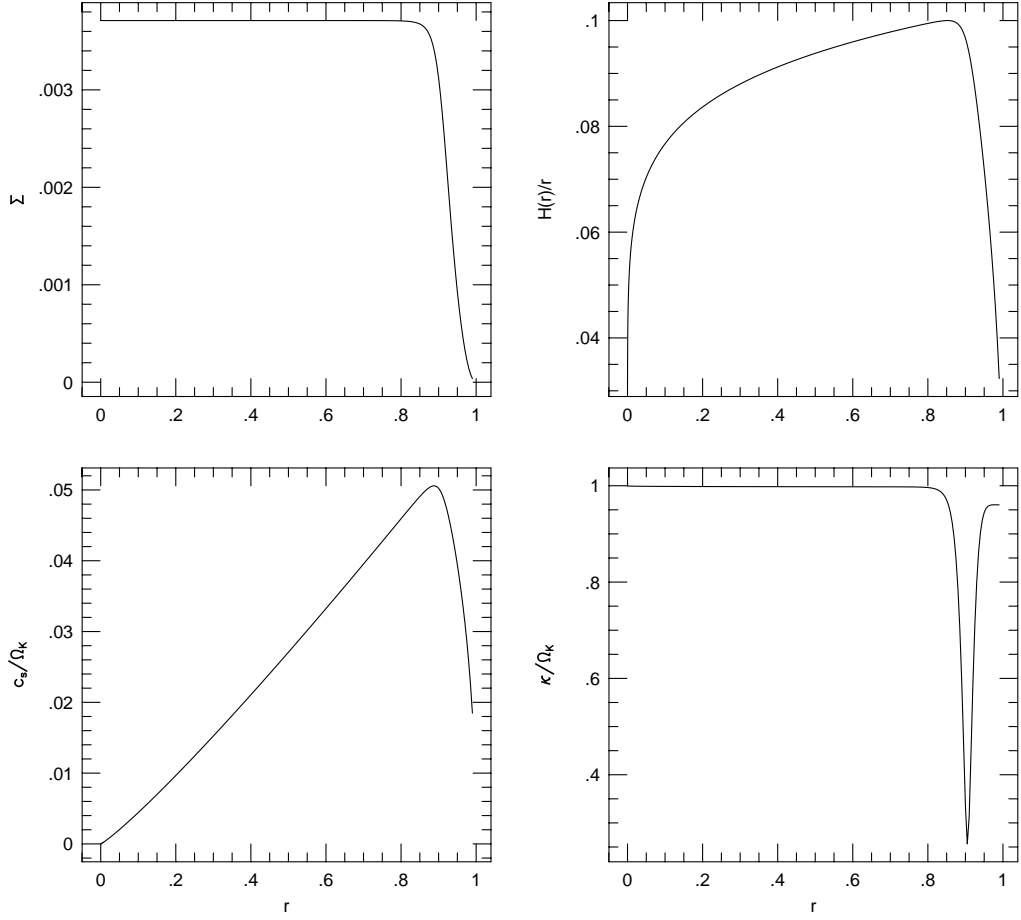}
\caption{ Surface mass density ({\it top left}), relative
semithickness ({\it top right}), ratio of the sound speed to the
Keplerian frequency ({\it bottom left}) and ratio of the epicyclic
frequency to the Keplerian frequency ({\it bottom right}) vs. $r$
for the equilibrium disk models. The parameters are $n=1.5,$
$M=10^{-2}M_p$ and $(H/r)_{max}=0.1$.  }
\label{figa}
\end{figure}

Figure~\ref{figa} shows the surface mass density, $\Sigma,$ the 
relative semi--thickness, $H/r$, the ratio of the sound speed to the 
Keplerian frequency, $c_s/\Omega_K \sim H$, and the ratio of the 
epicyclic frequency to the Keplerian frequency, $\kappa/\Omega_K$, 
versus the radius $r$ for the equilibrium disk models described above 
with $M=10^{-2}M_p$ and $(H/r)_{max}=0.1$. The epicyclic frequency 
departs from $\Omega_K$ at the outer edge of the disk because of the 
abrupt decrease of $\Sigma$ there. 
 
\section{The disk response} 
\label{sec:response} 
 
\subsection{Linear Perturbations} 

In this paper we consider small perturbations, so that the basic
equations~(\ref{continuity}) and~(\ref{motion}) can be linearized.  As
the relevant linearization of the equations has already been discussed
by Papaloizou~\& Lin~(1995), we provide only an abbreviated discussion
here.  We denote Eulerian perturbations with a prime.  Because the
perturbing potential, $\Psi',$ is proportional to $exp \left[ i \left(
\varphi - \Omega_pt \right) \right],$ the $\varphi$ and $t$ dependence
of the induced perturbations is the same.  We denote with a tilde the
part of the Eulerian perturbations that depends only on $r$ and $z$.
For instance, the Eulerian perturbed radial velocity is denoted
$v'_r$, and we have $v'_r (r,\varphi,z,t) = \tilde{v}'_r (r,z) exp
\left[ i \left( \varphi - \Omega_pt \right) \right]$.

We define a quantity $g$ (the physical meaning of which will be given
at the end of this section) through:
 
\be 
\frac{P'}{\rho} + \Psi' = -irz \Omega^2 g ,
\label{g} 
\ee 

\noindent and we also have $g(r,\varphi,z,t)=\tilde{g} (r,z) exp
\left[ i \left( \varphi - \Omega_pt \right) \right]$.
 
It was first shown by Papaloizou~\& Pringle~(1983) that horizontal
motions induced in a near Keplerian tilted disk are resonantly
driven. The main effect of a small viscosity is then to act on the
vertical shear in the horizontal perturbed velocities $\partial
v'_r/\partial z$ and $\partial v'_{\varphi}/\partial z.$ Demianski~\&
Ivanov~(1997) and Ivanov~\& Illarionov~(1997) have recently given the
relativistic generalization of this effect.  Following Papaloizou~\&
Lin~(1995), we then retain only the horizontal components of the
viscous force perturbation:
 
\begin{equation}
F'_{\nu r} = \frac{\partial}{\partial z} \left( \rho
\nu_{\rm v} \frac{\partial v'_r}{\partial z} \right), \; \;
F'_{\nu \varphi} = \frac{\partial}{\partial z} \left(
\rho \nu_{\rm v} \frac{\partial v'_{\varphi}}{\partial z}
\right),
\label{Fv}
\end{equation} 
 
\noindent where $\nu_{\rm v}$ is the kinematic viscosity that couples
to the vertical shear. This approximation is accurate when
$\alpha_{\rm v} \ll 1$.  The numerical calculations have actually
shown that, when $\alpha_{\rm v}$ becomes larger than 0.1, the terms
that are neglected in this approximation become comparable to the
terms that are retained.  For this reason, we shall limit our
numerical calculations to values of $\alpha_{\rm v}$ smaller than 0.1.

Here we have assumed that the disk is turbulent and that dissipation
can be modeled as a turbulent viscosity. The most likely mechanism for
producing such a turbulent viscosity is the magnetorotational
Balbus--Hawley instability (Balbus~\& Hawley~1991). We note however
that this instability may not operate in all the parts of protostellar
disks (Balbus~\& Hawley~1998).  Also, it is not clear that the tidal
waves would be damped by the turbulent viscosity in the way described
by equations~(\ref{Fv}).  In this formulation, we implicitly assume
that the turbulent viscosity is not affected by the tidally--produced
velocities, which is not necessarily true.  However, since the
wavelength of the tidal waves we consider is much larger than the disk
semi--thickness, this formulation may be correct.
 
Papaloizou~\& Lin~(1995) have shown that when the radial wavelength of
the perturbations is larger than the disk semi--thickness, a vertical
averaging approximation, in which $\tilde{g}$ is assumed to be
independent of $z$, can be used. Each of $\tilde{v}'_r$ and
$\tilde{v}'_{\varphi}$ are then proportional to $z$, with
$\tilde{v}'_z$ being independent of $z$. They have found some support
for the procedure in a numerical study of the propagation of
disturbances in disk models that take the vertical structure fully into
account.
 
Linearization of the equation of motion followed by multiplication by
$\rho z$ and vertical averaging then gives the velocity perturbations
in the form:
 
\begin{displaymath}
\frac{\tilde{v}'_r}{\Omega z} = - \left( 1 - \Omega_p/\Omega
\right) \tilde{g} 
\end{displaymath}
\be
+ \frac{\left( 1 - \Omega_p/\Omega \right) r \frac{d
\tilde{g}}{dr} + \tilde{g} \left( 3- \Omega_p/\Omega \right)
\Omega_p^2/\Omega^2 } {\left( 1 - \Omega_p/\Omega - i \alpha_{\rm v}
\right)^2 - \kappa^2/\Omega^2} , \label{vr} \ee
 
\be \frac{\tilde{v}'_{\varphi}}{\Omega z} = \frac{i}{1 -
\Omega_p/\Omega} \left( \tilde{g} + \frac{\kappa^2}{2\Omega^2}
\frac{\tilde{v}'_r}{\Omega z} \right) , \label{vphi} \ee
 
\be \frac{\tilde{v}'_z}{r \Omega}= \frac{\tilde{g}}{1-\Omega_p/\Omega}
, \label{vz} \ee
 
\noindent where $\alpha_{\rm v}$ is a vertical average of the standard
Shakura~\& Sunyaev~(1973) parameter that couples to the vertical
shear. It is defined by:

\begin{equation}
\alpha_{\rm v} = \int_{- \infty}^{\infty} \rho \nu_{\rm v} dz \left/
\left( \Omega \int_{- \infty}^{\infty} \rho z^2 dz \right) \right. .
\label{alphav}
\end{equation}

As mentioned above, the main effect of viscosity is to damp the
resonantly driven horizontal velocities.  Therefore, the viscous terms
which do not act directly on the resonance have been neglected.  This
is consistent with keeping only the components of the viscous force
given by equation~(\ref{Fv}).  Again, this is accurate only when
$\alpha_{\rm v} \ll 1$.
 
The same procedure applied to the continuity equation gives:

\begin{displaymath} 
-i \left( \Omega-\Omega_p \right) {\cal{I}} + \frac{\Omega^2
\tilde{g}}{\left( \Omega-\Omega_p \right)} \left\{ r \left[ \left(
\Omega-\Omega_p \right)^2-\Omega^2 \right] {\cal{J}} + \frac{d\mu}{dr}
\right\} 
\end{displaymath}
\be
= - \frac{1}{r} \left[ \frac{\partial}{\partial r} \left( r
\mu \frac{\tilde{v}'_r}{z} \right) + i \mu
\frac{\tilde{v}'_{\varphi}}{z} \right] , \label{lcont} \ee
 
\noindent in which 
 
\begin{displaymath} 
{\cal{J}} = \int_{- \infty}^{+ \infty} \frac{\rho z^2}{c^2} dz , \;
\mu = \int_{- \infty}^{+ \infty} \rho z^2 dz , \;
{\cal{I}} = \int_{- \infty}^{+ \infty} \frac{\rho z}{c^2}
{\tilde{\Psi}}' dz .
\end{displaymath} 
 
\noindent We note that equation~(\ref{lcont}) depends on the viscosity 
only through the perturbed velocities. 
 
To close the system of equations, we need a relation between
$\tilde{P}'$ and $\tilde{\rho}'$. For simplicity, we assume that the
equation of state is the same for the perturbed and unperturbed
flows.  Linearization of the equation of state~(\ref{polyt}) then gives
 
\begin{equation} 
\tilde{P}'=c_s^2 \tilde{\rho}' . \label{statep}
\end{equation} 
 
We now give a physical interpretation of $g$. Equation~(\ref{vz}) can
be used to express $\tilde{g}$ in terms of the vertical Lagrangian
displacement $\tilde{\xi}_z$:
 
\be
\tilde{g}=i \left( 1-\Omega_p/\Omega \right)^2 \frac{\tilde{\xi}_z}{r}
,
\ee
 
\noindent where again we have defined $\tilde{\xi}_z$ such that
$\xi_z(r,\varphi,z,t) = \tilde{\xi}_z(r,z) exp \left[ i \left( \varphi
- \Omega_pt \right) \right]$.

\noindent When $\Omega_p \ll \Omega$, the physical relative vertical 
displacement is then given by:
 
\begin{equation} 
Re\left( \frac{\xi_z}{r} \right) = Re(\tilde{g})
\sin\left(\varphi-\Omega_pt\right) + Im(\tilde{g})
\cos\left(\varphi-\Omega_pt\right) . \label{vert}
\end{equation} 
 
\noindent Thus, when $\Omega_p=0$, $Re(\tilde{g})$ and $Im(\tilde{g})$
are the relative vertical displacement along the $y$ and $x$--axis
respectively. We note that a constant value of $\tilde{g}$ corresponds
to a rigid tilt.
 
\subsection{Zero--Frequency Response} 
\label{sec:zero} 
 
From equation~(\ref{tPsi'}), we have to consider the disk response to
a secular potential perturbation with zero forcing frequency. When
$\Omega_p=0$ and $\alpha_{\rm v} \ll 1,$ equations~(\ref{vr}),
(\ref{vphi}), (\ref{vz}) and~(\ref{lcont}) reduce to the single
second--order ordinary differential equation for $\tilde{g}$
(Papaloizou~\& Lin~1995):
 
\be \frac{d}{dr} \left[ \frac{\mu}{\left( 1 - i \alpha_{\rm v}
\right)^2 - \kappa^2/\Omega^2} \frac{d \tilde{g}}{dr} \right] =
\frac{i {\cal{I}}}{r} .
\label{ODE} \ee 
 
\noindent Here we have assumed that:  
 
\be
\mu \ll Re \left\{ r \frac{d}{dr} \left[ \frac{\mu}{\left(
1-i\alpha_{\rm v} \right)^2 - \kappa^2/\Omega^2} \right] \right\} ,
\ee
 
\noindent which is certainly a reasonable approximation for
$\alpha_{\rm v} \ll 1$.
 
It can be shown that with the term $\cal{I}$ in its present form, the
$x$--component of the torque is non zero. This torque leads to the
precession of the disk, as in a gyroscope. By writing the equation of
motion for $\Omega_p=0$ in the frame defined in
\S~\ref{sec:potential}, we have assumed that the response of the disk
appears steady in a nonrotating frame. Because of the precessional
motion induced by the secular perturbation, the response is actually
steady in a frame rotating with the precession frequency. The Coriolis
force corresponding to this additional motion must then be added in
the equation of motion (the centrifugal force can be neglected since
it is of second order in the precession frequency). The magnitude of
the Coriolis force in the frame in which the response appears steady
has to be such that the $x$--component of the total torque (involving
both the Coriolis and the gravitational forces) is zero. 

This requirement leads to the expression of the precession frequency
$\omega_p$ (Papaloizou~\& Terquem~1995; see also Kuijken~1991 in the
context of galactic disks):

\begin{eqnarray}
\frac{\omega_p}{\Omega_0} = & - & \frac{3}{4} \frac{M_s}{M_s+M_p} 
\left( \frac{\omega}{\Omega_0} \right)^2 \cos \delta 
\nonumber \\ 
& \times & \int_{r_{in}}^R 
\frac{\Sigma}{\left(\Omega/\Omega_0\right)^2} dr \left/ 
\int_{r_{in}}^R \frac{\Sigma}{\Omega/\Omega_0} dr 
\right. , \label{prec} 
\end{eqnarray}
 
\noindent where $\Omega_0=\Omega(R)$ and $r_{in}$ is the disk inner
boundary. Here the disk has been assumed to precess as a rigid
body. This is expected if the disk can communicate with itself, either
through wave propagation or viscous diffusion (self--gravity being
ignored), on a timescale less than the precession period. As long as
$\alpha_{\rm v}$ is smaller than the disk aspect ratio $H/r$ (which is
almost constant throughout most of the disk), bending waves may
propagate (Papaloizou~\& Pringle~1983; Papaloizou~\& Lin~1995). The
ability of these waves to propagate throughout the disk during a
precession time implies approximately that $H/r > \left| \omega_p
\right| /\Omega_0$. This is the condition for near rigid body
precession in an inviscid disk (Papaloizou~\& Terquem~1995). The
numerical simulations of Larwood et al.~(1996) show that this
condition also holds in a viscous disk in which $\alpha_{\rm v}$ does
not exceed $H/r$. This indicates that in the absence of self--gravity,
as long as $\alpha_{\rm v} < H/r$ (which is likely to be satisfied in
protostellar disks), pressure is the factor that controls the
efficiency of communication throughout the disk and then the
precessional behavior of the disk. A more complete discussion of the
precession of warped disks (which includes self--gravity) is given by
Papaloizou et al. (1998).
 
The Coriolis force can be taken into account by replacing
$\tilde{\Psi}'$ with:
 
\be \tilde{\Psi}' + 2i \omega_p rz \Omega sin \delta . \label{Psi't}
\ee
 
\noindent For the zero--frequency case, $\tilde{\Psi}'$ will represent
the total term~(\ref{Psi't}) and the real amplitude $f$ will represent
$f+2 \omega_p \Omega \sin \delta$ from now on.
 
\noindent With this new definition of $f$, $\tilde{g}$ is then given
by:
 
\be \tilde{g} \left( r \right) = - \int_{r_{in}}^{r} \frac{ \left( 1 -
i \alpha_{\rm v} \right)^2 - \kappa^2/\Omega^2}{\mu}
\int_{r_{in}}^{r'} {\cal{J}} \left( r'' \right) f \left( r'' \right)
dr'' dr' ,
\label{zerog} \ee 
 
\noindent where the quantity in the outer integral has to be evaluated
at the radius $r'$. We have supposed that $\tilde{g}=0$ at $r=r_{in}$,
but we note that an arbitrary constant may be added to
$\tilde{g}$. This corresponds to an arbitrarily small rigid tilt that
we assume eliminated by the choice of coordinate system.

The above expression shows that, depending on whether $\alpha_{\rm v}$
is larger or smaller than $1-\kappa^2/\Omega^2 \sim H^2/r^2$, the
imaginary part of $\tilde{g}$ is larger or smaller than its real part,
respectively. According to equation~(\ref{vert}), this corresponds to
a vertical displacement produced by the secular response being mainly
along the $x$-- or $y$--axis, respectively.

We finally comment that the above treatment is accurate only when
$\left| \omega_p \right|/\Omega_0$ is smaller than the maximum of
$1-\kappa^2/\Omega^2 \sim H^2/r^2$ and $\alpha_{\rm v}$. If this is
not the case, then the precession frequency has to be taken into
account in the resonant denominator $\left( 1 - i \alpha_{\rm v}
\right)^2 - \kappa^2/\Omega^2$ where it would be the dominant term.
 
\subsection{Finite--Frequency Response} 
 
We now consider the response generated in the disk by those terms in
the perturbing potential with finite forcing frequency. When $\Omega_p
\ne 0$, the second--order differential equation for $\tilde{g}$
obtained from equations~(\ref{vr}), (\ref{vphi}), (\ref{vz})
and~(\ref{lcont}) takes on the general form:
 
\be A \frac{d^2 \tilde{g}}{dr^2} + B \frac{d \tilde{g}}{dr} + C
\tilde{g} = S , \label{ODEc} \ee
 
\noindent with  
 
\be
A= \frac{r \left( \Omega - \Omega_p \right)}{D_2} \mu , 
\ee
 
\begin{displaymath}
B= \mu \Omega \left\{ \frac{1}{D_2} \left[ \left(1 + r \frac{dln 
\mu}{dr} \right) D_1 - \frac{\kappa^2}{2 \Omega^2}+ \left( 3- 
\frac{\Omega_p}{\Omega} \right) \frac{\Omega_p^2}{\Omega^2} \right] 
\right.
\end{displaymath}
\be
\left. -D_1 +\frac{1}{\Omega} \frac{d}{dr} \left( \frac{D_1 r
\Omega}{D_2} \right) \right\} , \ee
 
\be
S = i \left( \Omega - \Omega_p \right) {\cal{I}} , 
\ee
 
\noindent where we have set $D_1=1-\Omega_p/\Omega$ and
$D_2=\left(1-\Omega_p/\Omega-i\alpha_{\rm v}
\right)^2-\kappa^2/\Omega^2$. By introducing the function $h$ such
that:
 
\be
\frac{1}{h} \frac{dh}{dr} = \frac{1}{A} \left( B - \frac{dA}{dr} 
\right) , 
\ee
 
\noindent equation~(\ref{ODEc}) can be recast in the form:
 
\be \frac{d}{dr} \left( h A \frac{d \tilde{g}}{dr} \right) + h C
\tilde{g} = h S , \ee
 
\noindent the solutions of which are given by the Green's function 
method: 
 
\begin{displaymath}
\tilde{g} \left( r \right) = \frac{1}{W} 
\left[ g_1 \left( r
\right) \int_{r_{in}}^r g_2 \left( r' \right) s \left( r' \right) dr'
\right.
\end{displaymath}
\be
\left.
+ g_2 \left( r \right) \int_r^R g_1 \left( r' \right) s \left( r'
\right) dr' \right] . \label{Gg} \ee
 
\noindent Here $s=h \left(\Omega-\Omega_p \right) f r \Sigma 
\Omega^{-2}$, $W$ is the constant 
 
\be
W = h A \left( g_1 \frac{dg_2}{dr} - \frac{dg_1}{dr} g_2 \right) , 
\ee
 
\noindent and $g_1$ and $g_2$ are the solutions of the homogeneous
differential equation that satisfy the outer and inner boundary
conditions respectively. In the numerical calculations presented
below, $g_1$ and $g_2$ are calculated directly from
equations~(\ref{vr}),~(\ref{vphi}) and~(\ref{lcont}) using the
fourth--order Runge--Kutta procedure with adaptative stepsize control
given by Press~et al.~(1986). Computation of the function $h$ then
allows $\tilde{g}$ to be calculated from~(\ref{Gg}).
 
\subsection{Boundary Conditions} 
 
Equation~(\ref{ODEc}) has a regular singularity at the outer edge 
$r=R$ (see Papaloizou~\& Terquem~1995). Our outer boundary 
condition is thus that the solution be regular there. 
 
We take the inner boundary to be perfectly reflective. The radial
velocity in spherical polar coordinates, $\left(r \tilde{v}'_r+z
\tilde{v}'_z \right) / \left(r^2 + z^2 \right)^{1/2}$, thus vanishes
at the locations $(r,z)$ such that $r^2+z^2=r_{in}^2$, where $r_{in}$
is the disk inner radius. Using the expression~(\ref{vz}) of
$\tilde{v}'_z$, this means that at these locations:
 
\be
rz \left[ \frac{\tilde{v}'_r}{z} + \frac{\Omega
\tilde{g}}{1-\Omega/\Omega_p} \right] = 0 ,
\ee
 
\noindent and then, by continuity: 
 
\begin{equation} 
\frac{\tilde{v}'_r}{z} + \frac{\Omega \tilde{g}}{1-\Omega/\Omega_p} =
0 ,
\end{equation} 
 
\noindent at the inner boundary. 
 
We note that for the largest values of $\alpha_{\rm v}$, we expect the
waves to be damped before reaching the center of the disk, such that
the results are independent of the inner boundary condition. For
$\alpha_{\rm v}=0$, since there is no dissipation at the boundaries,
we expect no torque to be exerted on the disk and no net angular
momentum flux to flow through its boundaries.
 
\subsection{Domain of Validity of This Analysis}
\label{sec:validity}

In this analysis, we have neglected the variation of $\tilde{g}$ with
$z$. As noted above, Papaloizou~\& Lin~(1995) have shown that this
approximation is valid when the scale of variation of $\tilde{g}$ with
radius is larger than the disk semi--thickness $H$, i.e. when

\begin{equation}
\left| \frac{1}{\tilde{g}} \frac{d \tilde{g}}{dr} \right| <
\frac{1}{H} .
\label{crit1}
\end{equation}

\noindent When this is not satisfied, dispersive effects, which have
been neglected here, should be taken into account.

Since we also consider linear waves, our analysis is valid as long as
the perturbed velocities are smaller than the sound speed. When they
become supersonic, shocks occur that damp the waves (Nelson~\&
Papaloizou~1998). However, it is not correct to compare the Eulerian
perturbed velocities given by the equations~(\ref{vr}), (\ref{vphi})
and~(\ref{vz}) with the sound speed to know whether the waves are
linear or not. This is because the system should be invariant under
the addition of a rigid tilt (i.e. a constant) to $\tilde{g}$. An
increase of $\tilde{g}$ corresponding to the addition of a rigid tilt
could make the Eulerian perturbed velocities larger than the sound
speed without the waves getting nonlinear. In principle, as pointed
out by Papaloizou~\& Lin~(1995), to get perturbed quantities that do
not depend on the addition of a rigid tilt (i.e. that depend only on
the radial gradient of $\tilde{g}$), we should use Lagrangian rather
than Eulerian variables. However, since the Eulerian perturbed radial
velocity in spherical polar coordinates depends only on the gradient
of $\tilde{g}$ in the limit of small perturbing frequency, we can
compare this quantity to the sound speed to decide whether the motion
is subsonic.  The criterion that has to be satisfied is then:

\begin{equation}
\left| \frac{rv'_r+zv'_z}{\left(r^2+z^2 \right)^{1/2} } \right| \simeq
\left| \frac{\Omega z r \frac{dg}{dr}} {\left( 1- \Omega_p/\Omega - i
\alpha_{\rm v} \right)^2 - \kappa^2 / \Omega^2} \right| < c_s ,
\label{crit2}
\end{equation}

\noindent where we have assumed that $\Omega_p \ll \Omega$ and $z \ll
r$. We have kept $\Omega_p/\Omega$ in the resonant denominator
because, even though this term is small compared with unity, it may
still be larger than $1-\kappa^2 / \Omega^2$.

The WKB dispersion relation for bending waves gives $\Omega_p = \left(
c_s / 2 \right) k $ where $k \sim 1/R$ is the wavenumber
(Papaloizou~\& Lin~1995). For finite perturbing frequency, we can then
rewrite the above criterion in the form:

\begin{equation}
\delta \tilde{g} < {\rm max} \left( \frac{H}{r}, \alpha_{\rm v}
\right) ,
\end{equation}

\noindent where $\delta \tilde{g}$ is the variation of the tilt angle
across the disk and we have used the fact that $1- \kappa^2 / \Omega^2
\sim H^2/r^2 < H/r$. By $H/r$ we mean the aspect ratio anywhere in the
disk, since this quantity does not vary significantly throughout the
disk. For $\Omega_p=0$, the condition we get is:

\begin{equation}
\delta \tilde{g} < {\rm max} \left( \frac{H^2}{r^2}, \alpha_{\rm v}
\right)
\end{equation}

\noindent As pointed out in \S~\ref{sec:zero}, depending on whether
$\alpha_{\rm v}$ is larger or smaller than $1-\kappa^2/\Omega^2$, the
imaginary part of $\tilde{g}$ is larger or smaller than its real part,
respectively. Therefore, the above criterion and equation~(\ref{vert})
tell us that the variation across the disk of the relative vertical
displacement produced by the secular perturbation is limited by
$\alpha_{\rm v}$ along the $x$--axis and by $H^2/r^2$ along the
$y$--axis.

\noindent We note that in an inviscid disk, the zero--frequency
perturbation produces $\delta \tilde{g} \sim \left| \omega_p
\right|/\Omega(R)$ (Papaloizou, Korycansky~\& Terquem~1995). Since, as
we have pointed out in \S~\ref{sec:zero}, our analysis is valid only
when $\left| \omega_p \right|/\Omega(R)$ is smaller than the maximum
of $H^2/r^2$ and $\alpha_{\rm v}$, the above criterion is always
satisfied in an inviscid disk.

The most likely situation in a protostellar disk corresponds to
$\alpha_{\rm v} < H/r$.  In that case, when the perturbed velocities
are close to the sound speed, the secular perturbation produces a
steady (in the precessing frame) tilt the variation of which across
the disk is on the order of $H^2/r^2$ or $\alpha_{\rm v}$, whichever
term is the largest.  Superposed on this steady tilt, bending waves
produced by the finite frequency terms propagate, corresponding to a
tilt the variation of which across the disk is on the order of
$H/r$. This is exactly what was observed in the SPH simulations
performed by Larwood~\& Papaloizou~(1997), in which $\alpha_{\rm v}$
was larger than $H^2/r^2$. We note that since in the case of
protostellar disks $H/r$ is likely to be close to 0.1, the variation
of the vertical displacement across the disk due to the nonsecular
tidal perturbations can be as large as about a tenth of the disk
radius while the perturbation remains linear.

\section{Angular momentum exchange} 
\label{sec:momentum} 
 
Tidal perturbation of a disk may lead to an exchange of angular
momentum between the disk and the perturber.
 
If the disk is inviscid and does not contain any corotation resonance,
the nature of the boundaries would determine whether such an exchange
takes place or not (see Lin~\& Papaloizou~1993 for example). Because
of the conservation of wave action in an inviscid disk, waves excited
at the outer boundary propagate through the disk with an increasing
amplitude if the disk surface density increases inward or is uniform
(see Lightill~1978 for example). It is usually assumed that they
become non linear before reaching the center and are dissipated
through interaction with the background flow. Thus, the inner boundary
can be taken to be dissipative (Papaloizou~\& Terquem~1995).
Dissipation at the boundary then introduces a phase lag between the
perturber and the disk response, enabling a net torque to be exerted
by the perturber. This torque is transferred to the disk through
dissipation of the waves at the boundary. Because of the conservation
of angular momentum, the net torque is equal to the difference of
angular momentum flux through the disk boundaries. This flux is
constant (independent of $r$) inside the disk, since there is not
dissipation there.
 
The situation is different when a corotation resonance is present in
the disk, since this singularity provides a location where angular
momentum can be absorbed or emitted (Goldreich~\& Tremaine~1979;
Goldreich~\& Nicholson~1989).
 
When the disk is viscous, its response is not in phase with the
perturber. A net torque is then exerted on the disk even if the
boundaries are reflective, and the angular momentum flux inside the
disk is not constant, since the perturbed velocities are viscously
dissipated. This situation has been investigated by Papaloizou~\&
Pringle~(1977) and Papaloizou~\& Lin~(1984) in the context of coplanar
orbit. Here we present the analysis corresponding to a non coplanar
orbit.
 
We note that whenever the perturber rotates outside the disk, the
torque exerted on the disk is negative. Through dissipation of the
waves, the disk then loses angular momentum.  We first derive an
expression of the torque in terms of $\tilde{g}$, and then relate it
to $\alpha_{\rm v}$ using angular momentum conservation.
 
\subsection{Expression for the Torque} 
 
The net torque exerted by the perturber on the disk is given by: 
 
\begin{displaymath} 
{\bf T} = - \int_V Re \left[ \rho + \tilde{\rho}'
e^{i\left(\varphi-\Omega_pt\right)} \right] {\bf{r}} 
\end{displaymath}
\be
\timesb Re \left[
\nab \left( \tilde{\Psi}' e^{i\left(\varphi-\Omega_pt\right)} \right)
\right] dV , \ee
 
\noindent where the integration is over the volume $V$ of the 
unperturbed disk. Because of the $\varphi$--periodicity, the 
first--order term is zero, and the $z$--component of the torque is 
then 
 
\be T_z = - \int_V Re \left[ \tilde{\rho}'
e^{i\left(\varphi-\Omega_pt\right)} \right] Re \left[
\frac{\partial}{\partial \varphi} \left( \tilde{\Psi}'
e^{i\left(\varphi-\Omega_pt\right)} \right) \right] dV ,
\label{Tz} \ee 
 
\noindent which is equivalent to  
 
\be T_z = \frac{1}{2} Im \left( \int_V \tilde{\rho}'^* \tilde{\Psi}'
dV \right) .
\label{Tzim} \ee 
 
\noindent Using equations~(\ref{g}) and~(\ref{statep}),
$\tilde{\rho}'$ in the above equation can be replaced in terms of
$\tilde{g}$. The expression of $T_z$ then involves ${\cal{J}}$, which,
in a Keplerian disk, is equal to $\Sigma \Omega^{-2}$. We then get
 
\be T_z = \pi \int_{r_{in}}^R Im \left( \tilde{g} \right) f \Sigma r^3
dr
\label{Tzd} \ee 
 
\noindent where $f$ is the real amplitude defined by~(\ref{f}) (which 
has to be modified when $\Omega_p=0$ in order to take into account the 
disk precession, see \S~\ref{sec:zero}). In the numerical calculations 
presented below, the torque $T_z$ will be computed from this 
expression. 
 
If $J_D$ is the total angular momentum of the disk, the tidal effects 
we have considered would remove the angular momentum content of the 
disk on a timescale 
 
\be t_d = \frac{J_D}{T_z} \sim \frac{4}{5} \frac{M \sqrt{GM_pR}}{T_z} 
. 
\label{td} \ee 
 
\noindent We shall compare this timescale with the viscous timescale
$t_{\nu, {\rm h}} \sim (r/H)^2 \Omega(R)^{-1}/\alpha_{\rm h}$ where we
will take $H/r=(H/r)_{max}$ (for our disk models, $H/r$ does not vary
by more than 20\% in the main body of the disk, as indicated in
Fig.~\ref{figa}). Here $\alpha_{\rm h}$ is the viscous parameter that
couples to the horizontal shear.
 
We also define the torque integral $T_z(r)$, which is the torque 
exerted between the inner boundary and the radius $r$: 
 
\be T_z(r) = \pi \int_{r_{in}}^{r} Im \left( \tilde{g} \right) f
\Sigma r^3 dr . \label{Tzdr} \ee
 
\subsection{Angular Momentum Conservation} 
\label{sec:angcon} 
 
To get the angular momentum conservation equation, which is of second 
order in the perturbed quantities, we first take the Lagrangian 
perturbation of the equation of motion~(\ref{motion}) and multiply by 
the unperturbed mass density. To first order, this leads to 
(Lynden--Bell~\& Ostriker~1967): 
 
\begin{displaymath}  
\rho \left[ \frac{\partial^2 \xib}{\partial t^2} + \Omega^2
\frac{\partial^2 \xib}{\partial \varphi^2} + \omegab \timesb \left
( \omegab \timesb \xib \right) + 2 \Omega \frac{\partial^2
\xib}{\partial \varphi \partial t} + 2 \omegab \timesb \frac{\partial
\xib}{\partial t} \right.
\end{displaymath}
\be
\left. + 2 \Omega \omegab \timesb \frac{\partial
\xib}{\partial \varphi} \right] = \nonumber \\ \rho \Delta \left
( -\frac{1}{\rho} \nab P \right) + {\bf{F'}}_{\nu} - \rho \nab \Psi' -
\rho \left( \xib \cdot \nab \right) \nab \Psi .
\label{ang} 
\ee
 
\noindent Here, $\xib$ is the Lagrangian displacement vector, $\Delta$
denotes the Lagrangian change operator, and we have used the relation
between Lagrangian and Eulerian perturbations of a quantity $Q$ to
first order in $\xib$:
 
\be
\Delta Q = Q' + \left( \xib \cdot \nab \right) Q . 
\ee
 
\noindent The vector $\omegab \equiv \Omega \veck$, with $\veck$ being
the unit vector in the $z$--direction. To derive equation~(\ref{ang}),
we have used the fact that at equilibrium ${\bf F}_{\nu}={\bf 0}.$
 
We now take the scalar product of equation~(\ref{ang}) with $\xibc,$
which is the complex conjugate of $\xib.$ Since $\partial/\partial t =
-i \Omega_p$ and $\partial/\partial \varphi = i,$ all the terms on the
left hand side of the resulting equation are real. We then take the
imaginary part of this equation and integrate it over the volume V of
the unperturbed disk. This leads to:
 
\begin{displaymath}
Im \left\{ \int_V \left[ \xibc \cdot {\cal{P}} \left( \xib \right)
+ \xibc \cdot {\bf{F'}}_{\nu} 
\right. \right.
\end{displaymath}
\be \left. \left.
- \rho \xibc \cdot \nab \Psi' -\rho
\xibc \cdot \left( \xib \cdot \nab \right) \nab \Psi \right] dV
\right\} = 0 , \label{im} \ee
 
\noindent where we have defined the linear operator ${\cal{P}}$ such 
that ${\cal{P}} \left( \xib \right) = \rho \Delta \left( -\nab P /\rho 
\right).$ We note that if the terms on the left hand side of 
equation~(\ref{ang}) had non zero contribution to equation~(\ref{im}), 
it would be related to the total time derivative of the perturbed 
specific angular momentum of the disk. 
 
Using the perturbed equation of state~(\ref{statep}) in the form
$\Delta P/P= \left( 1+1/n \right) \Delta \rho/\rho,$ it can be shown
(Lynden--Bell~\& Ostriker~1967) that:
 
\be \int_V \xibc \cdot {\cal{P}} \left( \xib \right) dV = B + \int_V 
\xibc \cdot {\cal{O}} \left( \xib \right) dV , \label{opp} \ee 
 
\noindent where $\cal{O}$ is a self--adjoint linear operator, and $B$ 
is a boundary term: 
 
\be
B = \int_S \left[ \frac{1}{n} \rho \left( \nab \cdot \xib \right) \xibc 
+ P \left( \xibc \cdot \nab \right) \xib \right] \cdot {\bf n} \, dS . 
\ee
 
\noindent The integration is over the surface $S$ of the disk, and 
${\bf n}$ is the unit vector perpendicular to this surface, oriented 
toward the disk exterior. Since $P$ and $\rho$ vanish at $r=R$ and 
$z=\pm H$ and we consider models for which $r_{in} \simeq 0$, we have 
$B=0$. In addition, since $\cal{O}$ is self--adjoint, the integral on 
the right hand side of equation~(\ref{opp}) is real. The term 
involving $\cal{P}$ in equation~(\ref{im}) is then zero. 
 
Using the fact that $\partial^2 \Psi /\partial r \partial z =
\partial^2 \Psi /\partial z \partial r,$ it can also be shown that the
term involving $\nab \Psi$ is zero.
 
Since equation~(\ref{im}) has no $\varphi$ and $t$--dependence, we can
now use the tilded variables again.  We define $\xibt$ such that $\xib
( r, \varphi, z, t ) = \xibt(r,z) exp \left[ i \left( \varphi -
\Omega_pt \right) \right]$. We define $\xibct$ similarly. The term
involving $\tilde{\Psi}'$ in equation~(\ref{im}) can be written as
follows:
 
\begin{displaymath} 
Im \left( \int_V \rho \xibct \cdot \nab \tilde{\Psi}' dV \right) 
\end{displaymath}
\begin{equation}
= Im \left( \int_V \nab \cdot \left( \rho \xibct \tilde{\Psi}' \right) dV
\right) + Im \left( \int_V \tilde{\Psi}' \tilde{\rho}'^* dV \right) ,
\label{interm1} 
\end{equation}  
 
\noindent where an integration by parts has been performed and the 
linearized continuity equation:
 
\be
\tilde{\rho}' = - \nab \cdot \left( \rho \xibt \right) ,
\ee
 
\noindent (see, for example, Tassoul~1978) has been used. The first 
term on the right--hand side of equation~(\ref{interm1}) is zero 
because $\rho$ vanishes at $r=R$ and $z=\pm H$ and $r_{in} \simeq 
0$. Equation~(\ref{im}) then becomes: 
 
\be Im \left( \int_V \tilde{\rho}'^* \tilde{\Psi}' dV \right) = Im
\left( \int_V \xibct \cdot {\bf{\tilde{F}'}}_{\nu} dV \right)
\label{im2} . \ee
 
\noindent From equation~(\ref{Tzim}), we see that this is equivalent
to:
 
\be T_z = Im \left( \pi \int_{r_{in}}^R \int_{-H}^H \xibct \cdot
{\bf{\tilde{F}'}}_{\nu} r dr dz \right) . \label{im3} \ee
 
We now use the relation between the Eulerian velocity perturbation and 
the Lagrangian displacement: 
 
\be
{\bf{\tilde{v}'}} + \left( \xibt \cdot \nab \right) {\bf{v}} =
\frac{\partial \xibt}{\partial t} + \left( {\bf{v}} \cdot \nab \right)
\xibt
\ee
 
\noindent together with the expression~(\ref{Fv}) of
${\bf{\tilde{F}'}}_{\nu}$, the relation~(\ref{alphav}) and the fact
that $\rho \left( \pm H \right) = 0$ to transform equation~(\ref{im3})
into:
 
\begin{displaymath}
T_z = -\pi \int_{r_{in}}^R \mu \alpha_{\rm v} r \frac{ \left|
\tilde{v}'_r/z \right|^2 + \left |\tilde{v}'_{\varphi}/z
\right|^2}{1-\Omega_p/\Omega} dr 
\end{displaymath}
\be
- \pi Im \left[ \int_{r_{in}}^R \mu
\alpha_{\rm v} r \frac{\left( \tilde{v}'^*_r/z \right) \left(
\tilde{v}'_{\varphi}/z \right)}{\left( 1-\Omega_p/\Omega \right)^2}
\left( 1 - \frac{\kappa^2}{2 \Omega^2} \right) \right]
. \label{dissip} \ee
 
In the numerical calculations presented below, the torque will be 
computed from equation~(\ref{Tzd}) and compared with the result of the 
above equation. 
 
In a Keplerian disk, $\kappa \simeq \Omega$. Then, when $\Omega_p \ll
\Omega$ and $\alpha_{\rm v} \ll 1$, equation~(\ref{vphi}) is well
approximated by:
 
\be \frac{\tilde{v}'_{\varphi}}{z} \simeq i \frac{\kappa^2}{2
\Omega^2} \frac{\tilde{v}'_r}{z} , \label{vphiap} \ee
 
\noindent so that $T_z$ can be approximated by:
 
\be T_z \simeq - \frac{3 \pi}{2} \int_{r_{in}}^R \mu \alpha_{\rm v} r
\left| \frac{\tilde{v}'_r}{z} \right|^2 dr . \label{Tzapp} \ee
 
\noindent As expected, the torque is negative. Here we consider linear
perturbations such that the perturbed velocities remain smaller than
the sound speed.  We have pointed out in \S~\ref{sec:validity} that,
since the system should be invariant under the addition of a rigid
tilt to $\tilde{g}$, it is not in principle correct to compare the
Eulerian perturbed velocities to the sound speed to know whether the
waves are linear or not.  However, since the dominant term in the
expression~(\ref{vr}) of $\tilde{v}'_r$ is the term proportional to
the gradient of $\tilde{g}$, we can compare this velocity to the sound
speed to decide whether the motion is subsonic.  Therefore we can get
an estimate of the maximum value of the torque by setting
$|\tilde{v}'_r| \sim c_s$, $\mu \sim \rho H^3$ and dropping the
integration. This leads to a minimum value of $t_d$ which is equal to
$t_{\nu , {\rm h}} \times \alpha_{\rm h} / \alpha_{\rm v}$. Only if
$\alpha_{\rm v}$ were larger than $\alpha_{\rm h}$ could $t_d$ be
smaller than $t_{\nu , {\rm h}}$.
 
An other useful quantity for interpreting our results is the flux of 
vertical angular momentum in the radial direction at radius $r$: 
 
\begin{equation} 
J(r)= \int_{z=-H}^H \int_{\varphi=0}^{2\pi} \rho r^2 Re \left( 
v'_{\varphi} \right) Re \left( v'_r \right) d\varphi dz , 
\end{equation} 
 
\noindent which can also be written 
 
\begin{equation} 
J(r)= \pi \mu r^2 Re \left[ \frac{\tilde{v}'_{\varphi}}{z} \left(
\frac{\tilde{v}'_r}{z} \right)^* \right] .
\end{equation} 
 
The torque $T_z(r)$ exerted between the radii $r_{in}$ and $r$ results 
in a change of angular momentum of the disk. Part of this angular 
momentum is advected through the boundaries (this is $J(r)-J(r_{in}) 
\simeq J(r)$ since $r_{in}\simeq 0$), and part is dissipated between 
these radii. Thus the quantity $T_z(r) - J(r)$ is the amount of 
angular momentum deposited in the disk between the inner boundary and 
the radius $r$. As we shall see, this quantity is negative, which 
means that the disk loses angular momentum. 

In the numerical calculations presented below, the torque integral
will be computed from equation~(\ref{Tzdr}) and $J(r)$ will be
computed directly from the velocities using equations~(\ref{vr})
and~(\ref{vphi}).
 
\section{Numerical results} 
\label{sec:numerical} 
 
We compute the torque exerted on protostellar disks by a perturber
with an orbit that is inclined to the plane of the disk. For a
coplanar system, one expects the circumprimary disk to be truncated by
tidal effects in such a way that its radius is not greater than about
one--third of the separation of the system (Papaloizou~\&
Pringle~1977). Larwood~et al.~(1996) have shown that tidal truncation
is only marginally affected by the lack of coplanarity.  Thus in our
models we adopt $D \ge 3R$.  Since we consider binary mass ratio of
unity ($M_s=M_p$), the perturbing potential is small compared with the
potential of the central star, justifying the assumption of small
perturbation.
 
For computational convenience, we normalize the units such that
$M_p=1$, $R=1$, and $\Omega_K(R)=1$. Apart for the calculation of the
epicyclic frequency, the rotation law in the disk is taken to be
Keplerian (the departure from the Keplerian law, which occurs mainly
at the disk outer edge, does not exceed $\sim 3\%$). The inner
boundary of the disk is $r_{in}=10^{-4},$ and the outer boundary is
taken to be $r_{out}=0.99$ rather than 1 in order to avoid a zero
surface mass density at the outer edge.
 
We present here the results of some disk response calculations for an
inclination $\delta=\pi/4,$ a polytropic index $n=1.5$ and a disk mass
$M=10^{-2}$. Results for an arbitrary inclination $\delta<\pi/2$
(prograde orbit) or $\pi/2 < \delta < \pi$ (retrograde orbit) can be
obtained straightforwardly since $T_z$ (and then $1/t_d$)
corresponding to $\Omega_p=\pm 2\omega$ and 0 is proportional to
$(1\pm\cos\delta)^2 \sin^2\delta$ and $\sin^2 2\delta$,
respectively. As shown by equation~(\ref{prec}), the precession
frequency scales with $\cos\delta$. Since $g$ is independent of $M$
(see equations~[\ref{vr}], [\ref{vphi}], [\ref{vz}]
and~[\ref{lcont}]), expression~(\ref{Tzd}) shows that $T_z \propto
M$. The quantities $\omega_p$, $t_d$ and $t_{\nu , {\rm h}}$ are
independent of $M$.

\subsection{Uniform $\alpha_{\rm v}$} 
 
In table~\ref{table1} we summarize the results obtained for different
values of the parameters, which are the separation of the system $D$,
the viscous parameter $\alpha_{\rm v}$ and the maximum value of the
relative semi--thickness of the disk $(H/r)_{max}.$ The quantities we
compute are the torque $T_z$ associated with each of the perturbing
frequencies $2\omega$ (prograde term), $-2\omega$ (retrograde term)
and 0, the disk precession frequency $\omega_p$ and the ratio of the
tidal timescale $t_d$ to the viscous timescale $t_{\nu , {\rm h}}$. We
have taken here $\alpha_{\rm h} = \alpha_{\rm v}$. The two values of
the torque have been computed using equations~(\ref{Tzd})
and~(\ref{dissip}), the result corresponding to
equation~(\ref{dissip}) being in parentheses. The timescale $t_d$ is
computed from equation~(\ref{td}) with $T_z$ being the sum of the
contributions from each frequency.
 
We note that when either one of the criterions~(\ref{crit1})
or~(\ref{crit2}) is not satisfied (cases indicated by a footnote), it
is always close to the disk outer edge.  This is because the density
varies more rapidly there than in the other parts of the disk.  Also
the scale of variation of $\tilde{g}$ never becomes smaller than $H$
by less than a factor of a few.  Similarly, when the radial velocity
in spherical polar coordinates becomes larger than the sound speed, it
is only by a factor of a few.  We note that the results we present in
this case can still be used since the perturbed quantities can be
reduced by choosing appropriate scaling parameters, like the initial
inclination $\delta$ or the mass of the disk $M$.

In all cases the condition $\left| \omega_p \right| / \Omega(R) <
{\rm{max}} \left(H^2/r^2, \alpha_{\rm v} \right)$ is satisfied (see
\S~\ref{sec:zero}).  The condition for near rigid body precession,
$H/R \gg \left| \omega_p \right| /\Omega(R)$, is also always
satisfied.
 
We note that there is a good agreement between the values of the 
torque given by equations~(\ref{Tzd}) and~(\ref{dissip}). In addition, 
in the cases we computed, we found this latter equation to be very 
well approximated by~(\ref{Tzapp}). 
 
To simplify the discussion, we note $T_z^+,$ $T_z^-$ and $T_z^0$ the 
torque associated respectively with $\Omega_p=2\omega$, $-2\omega$ and 
0. Furthermore, we separate the $\delta$--dependence by writing 
$T_z^{\pm}=T_{\pm} (1\pm\cos\delta)^2\sin^2\delta$ and $T_z^0=T_0 
\sin^22\delta$ with $T_{\pm}$ and $T_0$ being independent of $\delta$. 
 
Since part of this section is devoted to the tilt $\tilde{g}$, we
recall that $\tilde{g}$ is related to the relative vertical
displacement through equation~(\ref{vert}). When $\Omega_p=0$,
$Re(\tilde{g})$ and $Im(\tilde{g})$ are the relative vertical
displacement along the $y$ and $x$--axis respectively.
 
\subsubsection{Finite--Frequency Response} 
 
\begin{figure}
\plotone{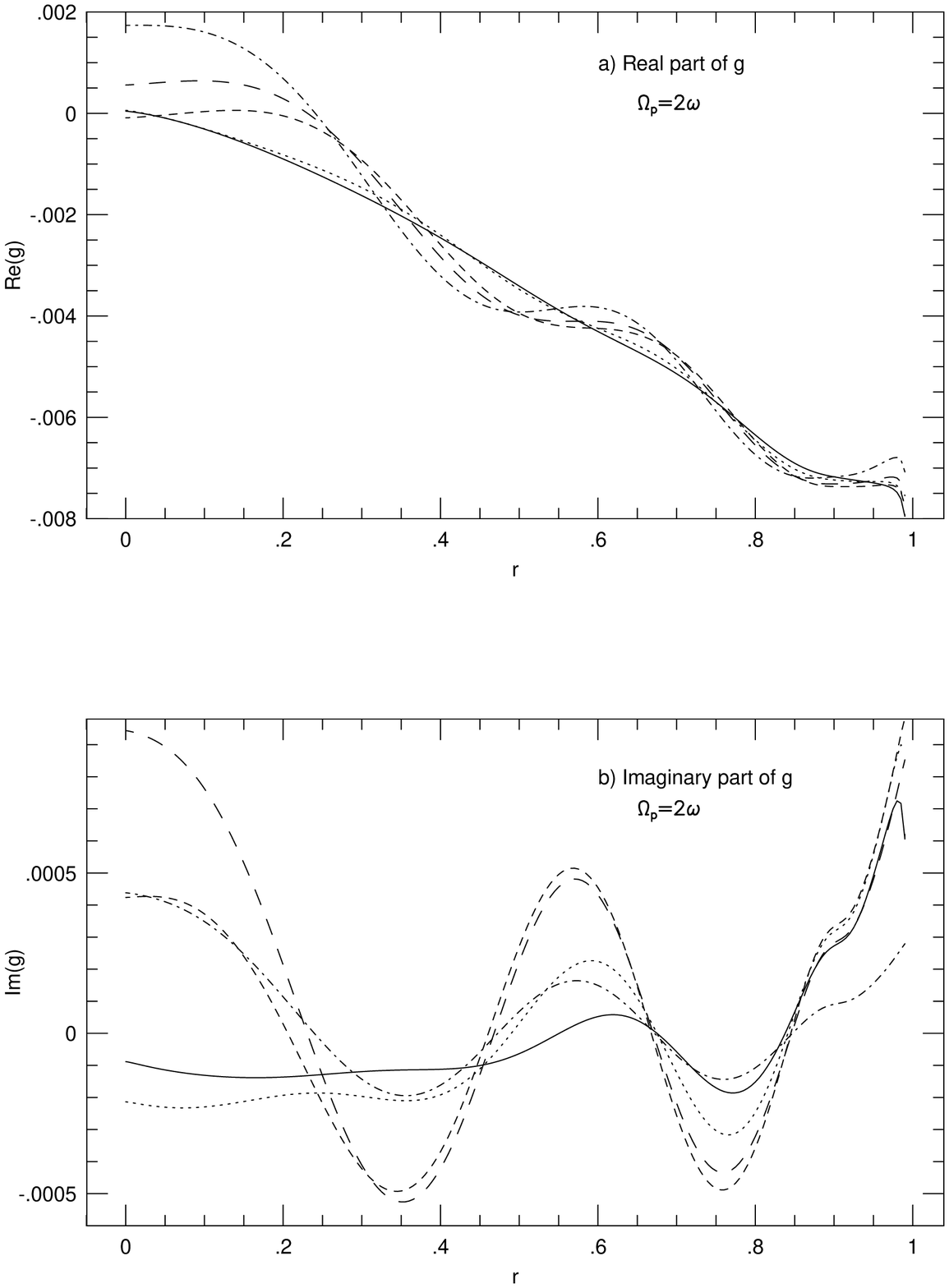}
\caption{ 
Real ({\it panel a}) and imaginary ({\it panel b}) parts of
$\tilde{g}$ vs. $r$ for models~2 and $\Omega_p=2\omega$. The
different curves correspond to $\alpha_{\rm v}=0.1$ ({\it solid
line}), $5 \times 10^{-2}$ ({\it dotted line}), $10^{-2}$ ({\it short
dashed line}), $5 \times 10^{-3}$ ({\it long dashed line}) and
$10^{-3}$ ({\it dot--short dashed line}). The parameters are $D=6$ and
$(H/r)_{max}=0.05$. The wavelike character of the response disappears
when $\alpha_{\rm v} \ge H/r.$
}
\label{fig1} 
\end{figure}

Figure~\ref{fig1} shows the real and imaginary parts of $\tilde{g}$
versus $r$ for models~2 and $\Omega_p=2\omega$ (figures for
$\Omega_p=-2\omega$ are similar). We see that the wave--like character
of the disk response disappears when $\alpha_{\rm v}$ becomes larger
than $H/r$ (which is 0.05 in these models).  This is in agreement with
Papaloizou~\& Pringle~(1983) and Papaloizou~\& Lin~(1995) (and also
with the relativistic generalization of Demianski~\& Ivanov~1997 and
Ivanov~\& Illarionov~1997) who have shown that, in a near Keplerian
disk (self--gravitating or not), the longest wavelength disturbances
undergo a transition between wave--like and diffusive behavior when
$\alpha_{\rm v} \sim H/r$.  We note, however, that when $\alpha_{\rm v}
\ge H/r$, there are still some oscillations in the half outer part of
the disk. This is because bending waves have a long wavelength, such
that they can still penetrate relatively far before being diffused
out.  \\ The magnitude of $Re(\tilde{g})$ does not vary with
$\alpha_{\rm v}$, since it is controlled mainly by the radial pressure
force. On the contrary, the magnitude of $Im(\tilde{g})$ is much more
sensitive to the viscosity, although it does not vary monotonically
with $\alpha_{\rm v}$.

By comparing models~2a and~3c and also models~2c and 4c, which differ
only by the value of $(H/r)_{max}$, we have checked that the wavelength
of the response is proportional to $H/r$. This is expected since
bending waves propagate with a velocity which, being half the sound
speed, is proportional to $H/r$, and the wavelength is proportional to
the wave velocity.
 
In all the cases we have computed, except model~3f, $T_z^+ \ge T_z^-$,
but $T_- \ge T_+$, in agreement with Papaloizou~\& Terquem~(1995).  In
general, the shorter the wavelength of the response, the smaller the
torque, and the wave corresponding to the retrograde term has a
slightly longer wavelength than that corresponding to the prograde
term.  The difference between $T_+$ and $T_-$ (and $T_0$) is still
significant for the largest values of the separation computed here.
We expect all these quantities to converge toward the same value when
$2\omega$ approaches 0, which probably means here
$2\omega/\Omega(R)\ll\alpha_{\rm v}$ since the torque is controlled by
viscosity.  This is not satisfied even for model~3i which has
$2\omega/\Omega(R)=0.09.$ However, we observe that the difference
between these torques is reduced from a factor 9 to a factor less than
4 when model~3i is run with $\alpha_{\rm v}=0.1$.
 
\begin{figure}
\plotone{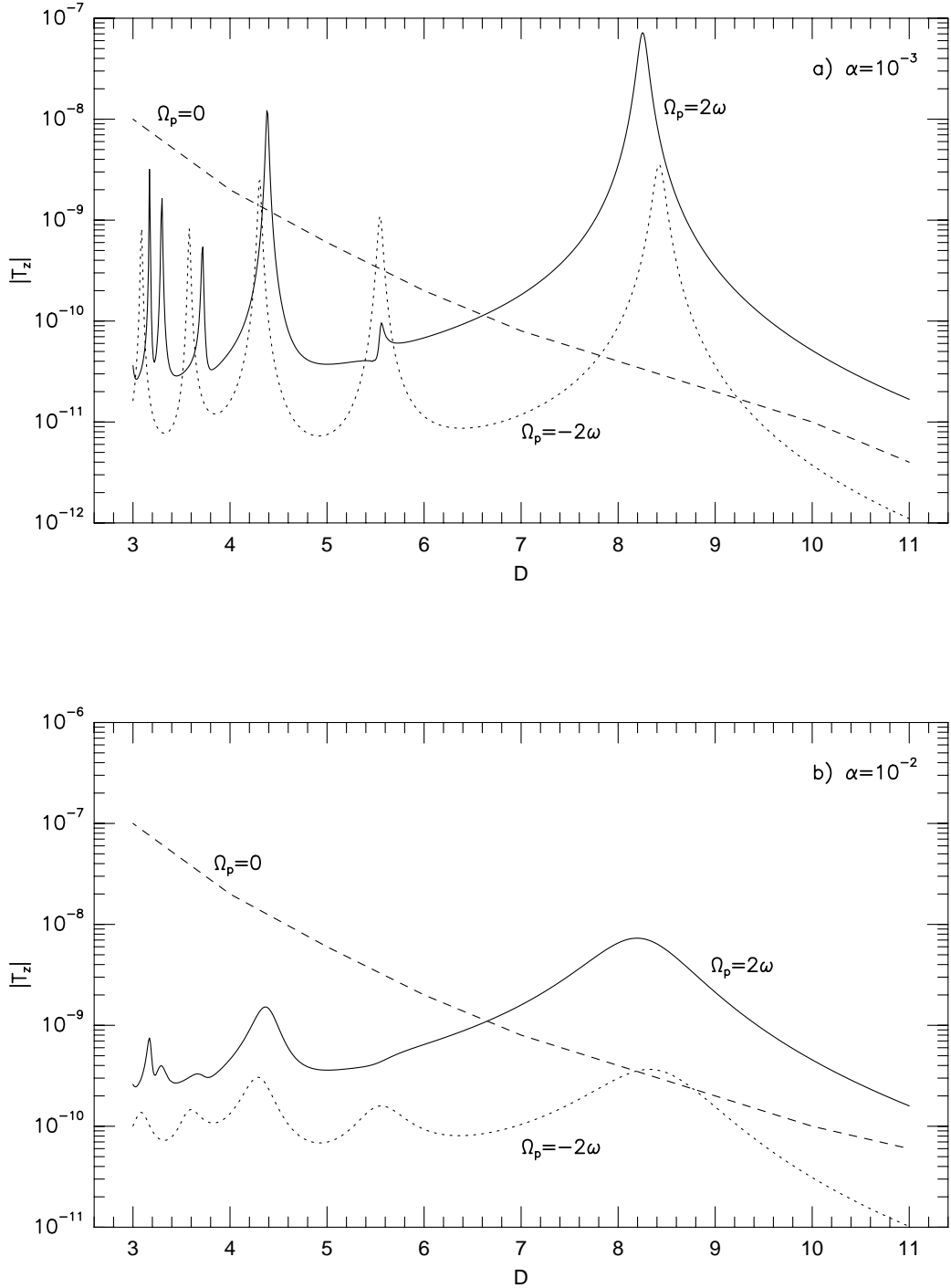}
\caption{ Net torque $\left|T_z\right|$ vs. $D$ in a semilog
representation for $(H/r)_{max}=0.1$. The panels {\it a} and {\it b}
correspond to $\alpha_{\rm v}=10^{-3}$ (models~1b and 3) and
$\alpha_{\rm v}=10^{-2}$ (models~1c and 4) respectively. The different
curves correspond to $\Omega_p=2\omega$ ({\it solid line}), $-2\omega$
({\it dotted line}) and 0 ({\it dashed line}). These plots show
several resonances.  }
\label{fig3} 
\end{figure}

In Figure~\ref{fig3} the net torque $T_z$ is plotted versus $D$ in a
semi--log representation for $\alpha_{\rm v}=10^{-3}$ (models~3 plus
other values of $D$) and $\alpha_{\rm v}=10^{-2}$ (models~4 plus other
values of $D$) and for $\Omega_p=\pm 2 \omega$ and 0.  Before
commenting on the resonances, we note that the torque associated with
the finite frequency terms out of resonance does not decrease when the
separation increases up to $D \sim 8$.  This confirms the results of
Papaloizou~\& Terquem~(1995).  There are competing effects acting when
the separation of the system is increased.  On the one hand, the
wavelength of the tidal waves becomes larger (since it is inversely
proportional to its frequency), which tends to increase the torque.
On the other hand, the amplitude of the perturbing potential
decreases, which tends to reduce the torque.  It seems that the first
effect is dominant for $D\le8$.  It seems like the torque begins to
decrease with the separation when the wavelength becomes comparable to
the disk radius, which happens for $D\sim 9.$ This is reasonable since
the wavelength cannot increase further.
 
In Figure~\ref{fig3} we can see resonances, which occur when the
frequency of the tidal waves is equal to that of some free normal
global bending mode of the disk, and cause the torque to become very
large, even infinite if there is no dissipation (Papaloizou~\&
Lin~1984). In a resonance, the torque indeed increases when
$\alpha_{\rm v}$ decreases, in contrast to what is observed out of
resonance.  It seems that there is no longer any resonance for
$D>10$.  This would not be surprising since the wavelength of the
response is then comparable to the disk radius.  A more detailed
description of the resonance which occurs at $D \sim 8.5$ is given in
the Appendix.

Since the resonances depend on the spectrum of the free normal bending
modes of the disk, they are model--dependent.  In particular, they
occur only if the the waves can be at least partially reflected at the
inner boundary, otherwise there is no free normal global bending mode
in the disk (see also Hunter~\& Toomre~1969 for the case of a purely
self--gravitating disk).
 
The values of the torque displayed in Figure~\ref{fig3} correspond to
$\delta = \pi/4$.  However, for this particular value of $\delta$, the
perturbed velocities may become larger than the sound speed close to a
resonance. These values of the torque should then be scaled to be used
for other values of $\delta$, or for other disk parameters.  For
$\alpha_{\rm v}=10^{-2}$, the resonances are very weak, indicating
that the waves are almost completely dissipated before reaching the
disk inner boundary.

We have compared our results with those obtained by Papaloizou~\&
Terquem~(1995) for an inviscid disk in which the inner boundary is
dissipative.  Remembering that they had $\delta=\pi/2$, we found that,
in general, we could reproduce their results with $\alpha_{\rm v} \sim
10^{-2}$.  This confirms that, when $\alpha_{\rm v} \sim 10^{-2}$, the
waves are dissipated before reaching the disk inner boundary.

Out of resonance, the torque $T_z^{\pm}$ associated with the finite
frequency terms is proportional to $\alpha_{\rm v}$ as long as this
parameter is smaller than some critical value. Then, for the larger
values of $\alpha_{\rm v}$, $T_z^{\pm}$ is independent of the
viscosity.  This is in agreement with the theoretical expectation
(Goldreich~\& Tremaine~1982) and the work of Papaloizou~\&
Terquem~(1995).  We indeed expect the torque to be independent of
whatever dissipation acts in the disk providing the waves are
dissipated before reaching the disk inner boundary.  If the waves are
reflected on the disk inner boundary, the situation is more
complicated because the angular momentum carried by the waves has a
different sign depending on whether the waves are in--going or
out--going.  Therefore part of the angular momentum lost by the disk
while the waves propagate inwards is given back by the waves
propagating outwards.  The net amount of angular momentum lost by the
disk then depends on how efficiently the waves are reflected.  Using
this argument, we then find that, when $H/r=0.1$ and 0.05, the waves
are damped before reaching the disk inner boundary when $\alpha_{\rm
v}$ is larger than $\sim 5 \times 10^{-2}$ and $\sim 5 \times
10^{-3}$, respectively (for $H/r=0.1$, this is a bit larger than the
value found above). Such a dependence of this critical value of
$\alpha_{\rm v}$ with $H/r$ is expected.  Indeed, the shorter the
wavelength, the more easily are the waves dissipated.  Since the
wavelength is proportional to $H/r$ (see above), we then expect that
for a given $\alpha_{\rm v}$ dissipation is more efficient for small
values of $H/r$.  For the same reason, the net torque $T_z^{\pm}$
increases with $H/r$ (see models~3c and~2a and models~4c and~2c).  The
coupling between the response and the perturbing potential is indeed
more efficient when the response has a long wavelength, and the
integral in equation~(\ref{Tzd}) from which the torque is calculated
is then larger.
 
\subsubsection{Zero--Frequency Response} 
 
\begin{figure}
\plotone{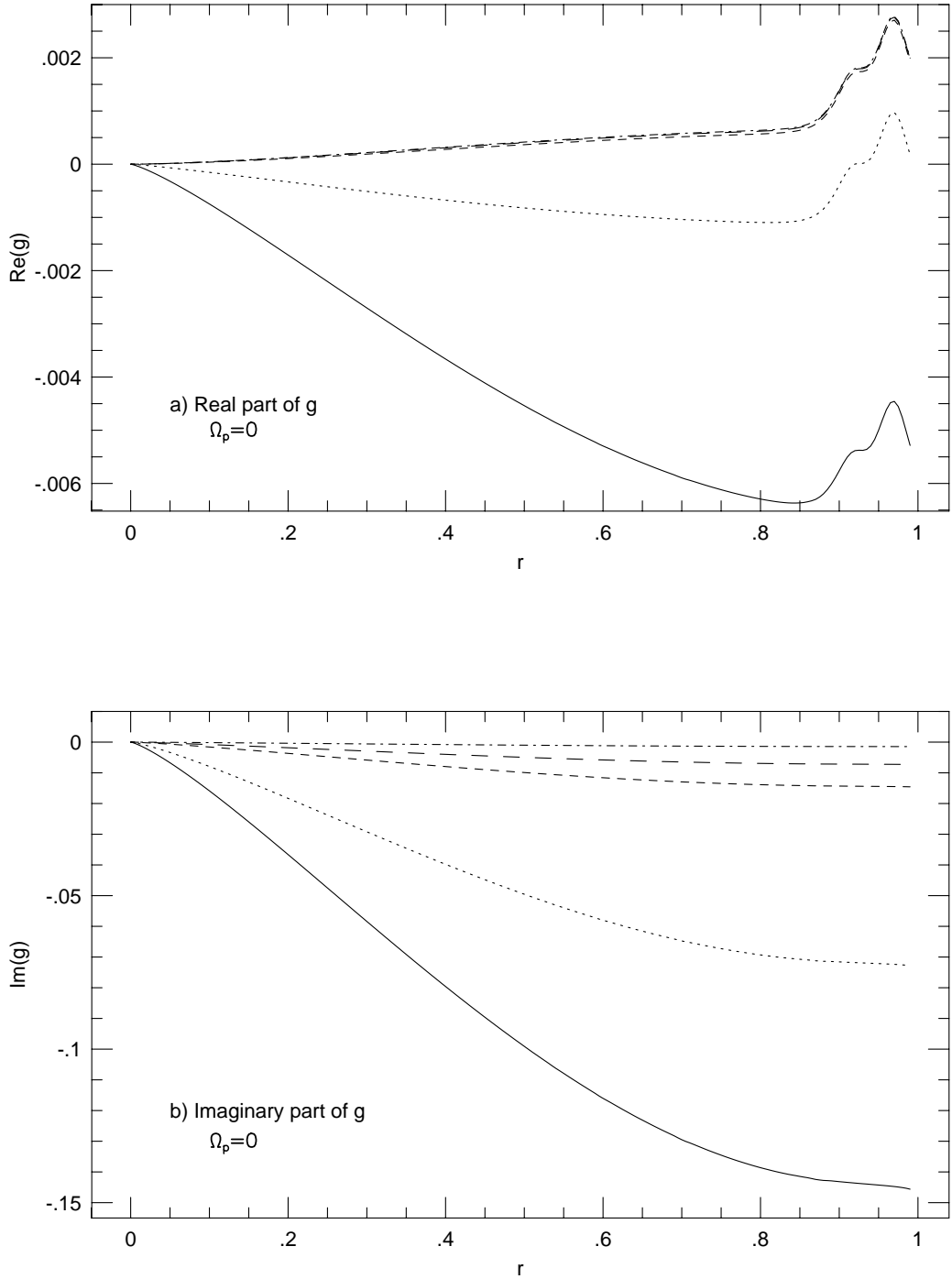}
\caption{
Real ({\it panel a}) and imaginary ({\it panel b}) parts of
$\tilde{g}$ vs. $r$ for models~2 and $\Omega_p=0$. The different
curves correspond to $\alpha_{\rm v}=0.1$ ({\it solid line}), $5
\times 10^{-2}$ ({\it dotted line}), $10^{-2}$ ({\it short dashed
line}), $5 \times 10^{-3}$ ({\it long dashed line}) and $10^{-3}$
({\it dot--short dashed line}). The parameters are $D=6$ and
$(H/r)_{max}=0.05$. The zero--frequency response is not wave--like.
}
\label{fig5} 
\end{figure}

Figure~\ref{fig5} shows the real and imaginary parts of $\tilde{g}$
versus $r$ for models~2 and $\Omega_p=0$.  As expected, we observe
that the zero--frequency response is not wave--like. 

From equation~(\ref{zerog}) we see that, as long as $\alpha_{\rm v}^2
\ll 1-\kappa^2/\Omega^2 \sim H^2/r^2$, which is the case for
$\alpha_{\rm v} \le 10^{-2}$, $Re(\tilde{g})$ is almost independent of
the viscosity.  When $\alpha_{\rm v}=0.1,$ $\alpha_{\rm v}^2$
dominates over $1-\kappa^2/\Omega^2$, and since it appears in the
expression of $Re(\tilde{g})$ with the minus sign, $Re(\tilde{g})$
becomes negative. The particular behavior of $Re(\tilde{g})$ in the
outer parts of the disk is produced by the fact that the surface mass
density drops to zero there, thus increasing $1-\kappa^2/\Omega^2$. As
expected from equation~(\ref{zerog}), $Im(\tilde{g})$ is proportional
to $\alpha_{\rm v}$.
 
Papaloizou, Korycansky~\& Terquem~(1995) computed the real part of
$\tilde{g}$ for an inviscid disk with an equilibrium state similar to
that we have set up here and for $\Omega_p=0$ and $D=7$. We have
checked that the results we get for the smallest values of
$\alpha_{\rm v}$ are in complete agreement with theirs.
 
Equation~(\ref{Tzd}) shows that $T_z$ depends only on $Im(\tilde{g})$,
not on $Re(\tilde{g})$.  Since for the secular response $Im(\tilde{g})
\propto \alpha_{\rm v}$, $T_z^0$ is also proportional to $\alpha_{\rm
v}$. This is borne out by the numerical results.  Also
equation~(\ref{zerog}) indicates that $\tilde{g} \propto 1/H^2.$ Thus
if we vary $H/r$ while keeping $\Sigma$ constant, we get from
equation~(\ref{Tzd}) that $T_z \propto 1/H^2$. This is confirmed by
models~3c and~2a and models~4c and~2c, which differ only by the value
of $(H/r)_{max}$. This is in contrast to the finite frequency response,
for which the torque increases with $H/r$ (see above).
 
Figure~\ref{fig3} indicates that $T_z^0 \gg T_z^{\pm}$ out of
resonance. Since we also have $T_0 \gg T_{\pm}$, $T_z^0$ is always
going to be much larger than $T_z^{\pm}$ for $\delta$ not too close to
$\pi/2$ (typically for $\delta<70^{\arcdeg}$ or
$110^{\arcdeg}<\delta<180^{\arcdeg}$).
 
In contrast to the finite frequency response, the zero--frequency
response is not wave--like, and we then expect $T_z^0$ to decrease
continuously with $D$.  This is indeed what we observe in
Figure~\ref{fig3}.
 
\subsubsection{Angular Momentum Dissipation as a Function of Radius} 
 
\begin{figure}
\plotone{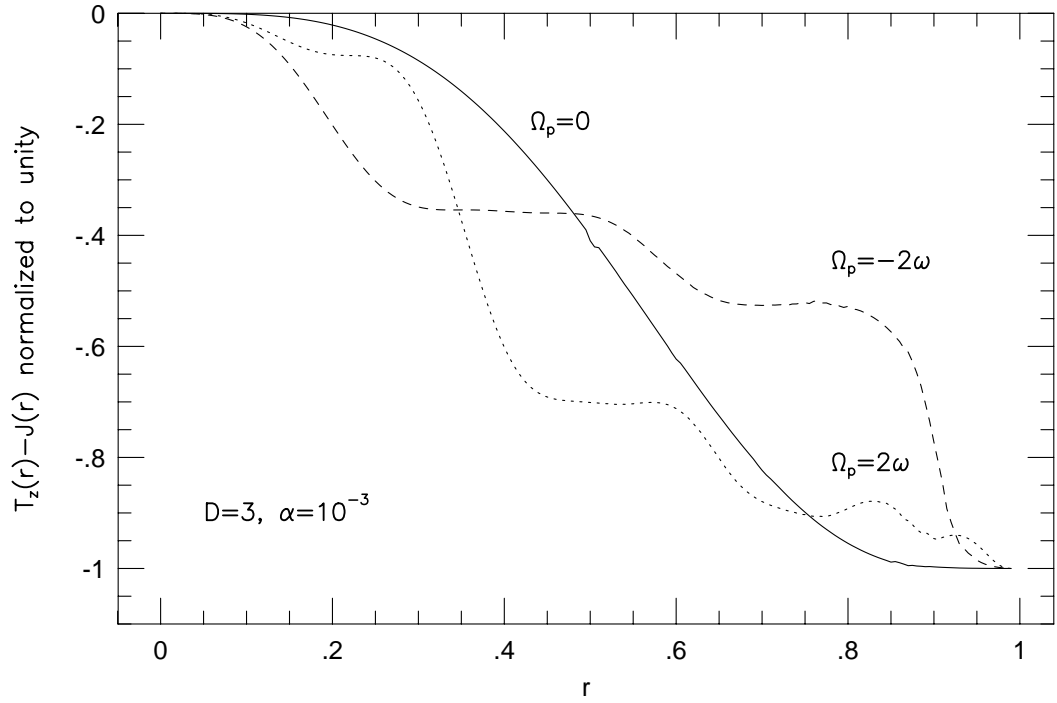}
\caption{
$T_z(r)-J(r)$ (angular momentum dissipated between $r_{in}$ and $r$)
normalized to unity vs. $r$ for $\alpha=10^{-3}$, $D=3$ (model~1b)
and $\Omega_p=0$ ({\it solid line}), $2\omega$ ({\it dotted line}) and
$-2 \omega$ ({\it dashed line}).
}
\label{fig3ter} 
\end{figure}

\begin{figure}
\plotone{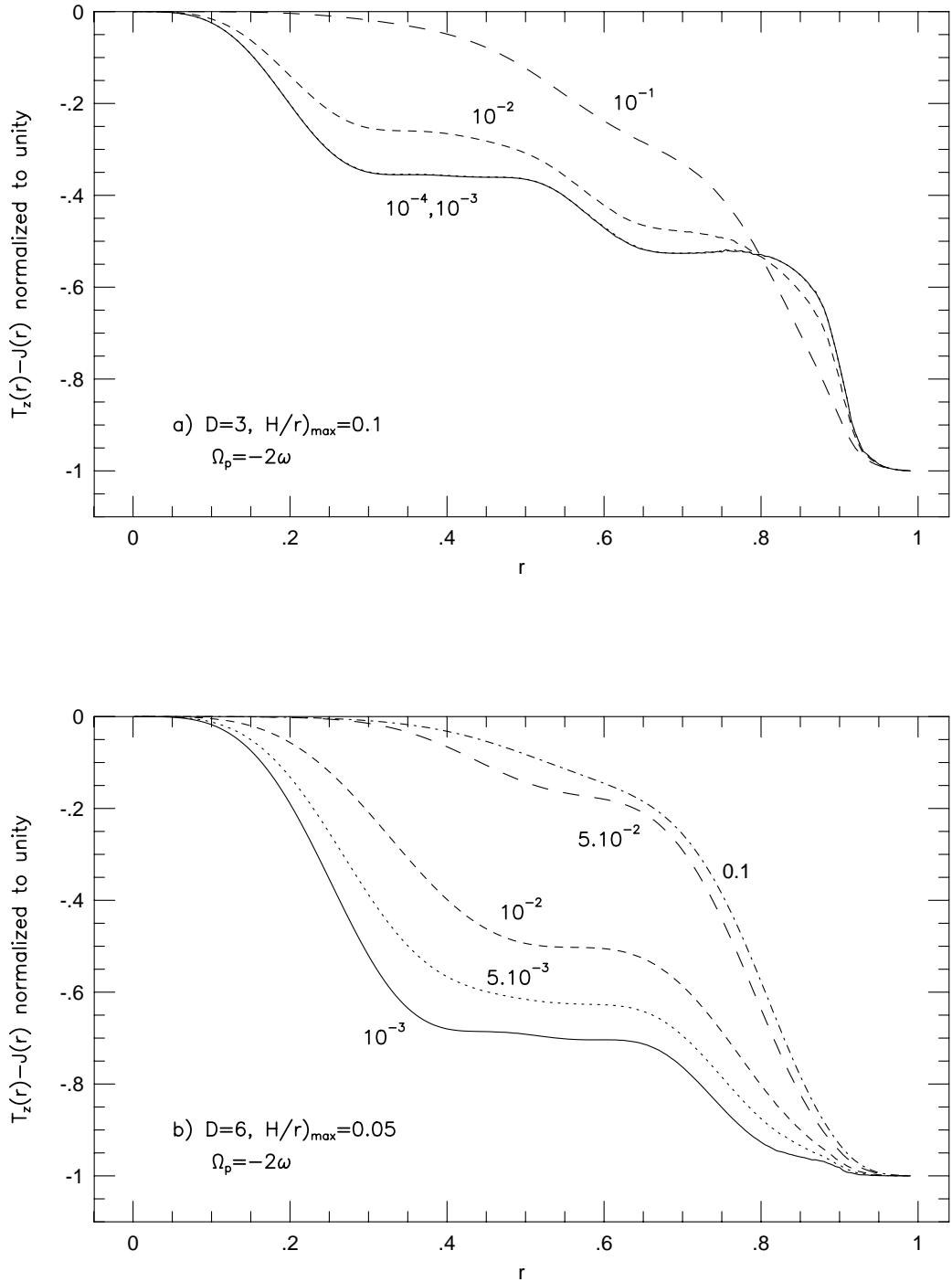}
\caption{
$T_z(r)-J(r)$ (angular momentum dissipated between $r_{in}$ and $r$)
normalized to unity vs. $r$ for $\Omega_p=-2\omega$. The panel (a)
corresponds to models~1, i.e. $D=3$, $(H/r)_{max}=0.1$ and
$\alpha=10^{-4}$ ({\it solid line}), $10^{-3}$ ({\it dotted line}),
$10^{-2}$ ({\it short dashed line}) and 0.1 ({\it long dashed
line}). The panel (b) corresponds to models~2, i.e. $D=6$,
$(H/r)_{max}=0.05$ and $\alpha=10^{-3}$ ({\it solid line}), $5 \times
10^{-3}$ ({\it dotted line}), $10^{-2}$ ({\it short dashed line}), $5
\times 10^{-2}$ ({\it long dashed line}) and 0.1 ({\it dot--short
dashed line}).
}
\label{fig3q} 
\end{figure}

Figures~\ref{fig3ter} and~\ref{fig3q} show $T_z(r)-J(r)$ normalized to 
unity versus $r$ for various models. This represents the angular 
momentum deposited in the disk between the inner boundary and the 
radius $r$. The fact that this quantity is negative means that the 
disk loses angular momentum.  Of course, $\left| T_z(r)-J(r) \right|$ 
should increase with $r$.  However, we observe in Figure~\ref{fig3ter} 
that for $D=3$ and $\Omega=2\omega$ this function has a little 'hump' 
close to the disk outer edge.  We interpret it as being a numerical 
effect, since its amplitude is found to depend on the accuracy of the 
calculations. 
 
We first observe that bending waves are able to transport a
significant fraction of the negative angular momentum they carry deep
into the inner parts of the disk.  For the parameters corresponding to
model~1b, Figure~\ref{fig3ter} shows that the retrograde and prograde
waves transport respectively 35\% and 60\% of the total negative
angular momentum they deposit into the disk at radii smaller than 0.3
and 0.4 respectively.  In contrast, only 20\% of the total negative
angular momentum carried by the zero--frequency perturbation is
deposited at radii smaller than 0.4.  We note however that, out of
resonance, the net amount of angular momentum deposited at small radii
by the $\Omega_p=0$ perturbation is not necessarily smaller than that
deposited by the finite frequency perturbations. Also, the fraction of
angular momentum deposited close to the disk outer edge by the finite
frequency responses (especially the retrograde wave) is larger than
that deposited by the secular response.

It appears that the secular response gets dissipated very
progressively throughout the disk, whereas the prograde and retrograde
waves are dissipated preferentially at some locations in the disk.
The position of these locations seems to depend on the wavelength of
these waves.
 

The resonances do not affect the distribution of dissipation of
angular momentum in the disk, although they obviously increase the
absolute value of the actual amount of angular momentum deposited.
This is due to the fact that the magnitude of the response, not its
wavelength, is affected by a resonance.
 
In Figure~\ref{fig3q} we have plotted $T_z(r)-J(r)$ versus $r$ for
$\Omega_p=-2\omega$ and different values of $\alpha_{\rm v}$. We see
clearly on these plots the transition that occurs when $\alpha_{\rm v}
\sim H/r$. When $\alpha_{\rm v}$ becomes larger than $H/r$, the
wave--like response of the disk disappears.  Thus, the curves
corresponding to $\Omega_p=-2\omega$ have a shape similar to those
corresponding to the secular response. 

In general, when $\alpha_{\rm v}$ is increased, the fraction of
negative angular momentum deposited at small radii decreases. This is
expected since the waves, excited predominantly in the outer parts of
the disk, get dissipated more easily at large radii when $\alpha_{\rm
v}$ is large. 

We observe in Figure~\ref{fig3q}.a that there is hardly any difference
between the cases $\alpha_{\rm v}=10^{-4}$ and $\alpha_{\rm
v}=10^{-3}$. This means that whatever the amount of angular momentum
transported by the wave, the same fraction is deposited between
$r_{in}$ and $r$ in both cases. This probably indicates that the wave
is able to reach the disk inner boundary for these values of
$\alpha_{\rm v}$. For larger $\alpha_{\rm v}$ for which the wave is
completely dissipated before reaching the inner boundary, no angular
momentum is deposited at small radii, so that the curves look
different.  This indicates that, for $(H/r)_{max}=0.1$, the waves do
not reach the inner boundary when $\alpha_{\rm v}$ is larger than
$\sim 10^{-2}$.  For $(H/r)_{max}=0.05$ (Figure~\ref{fig3q}.b), the
critical value of $\alpha_{\rm v} $ above which the waves do not reach
the inner boundary is between $10^{-3}$ and $5 \times 10^{-3}$.  These
values are in agreement with those found above.  Our calculations also
show that a larger fraction of the negative angular momentum is
deposited at small radii when $H/r$ is decreased.  However, we have
checked that this effect is less significant when the distance
interior to the outer boundary occupied by the taper of the surface
mass density is reduced along with the semi--thickness.
 
\subsection{Nonuniform $\alpha_{\rm v}$} 
 
We have run model~1c with a nonuniform viscosity $\alpha_{\rm
v}=10^{-2}x^5$ and $\alpha_{\rm v}=10^{-2} \left(1-x \right)^5$. These
models correspond to a disk where respectively the inner and outer
parts are nonviscous.
 
\begin{figure}
\plotone{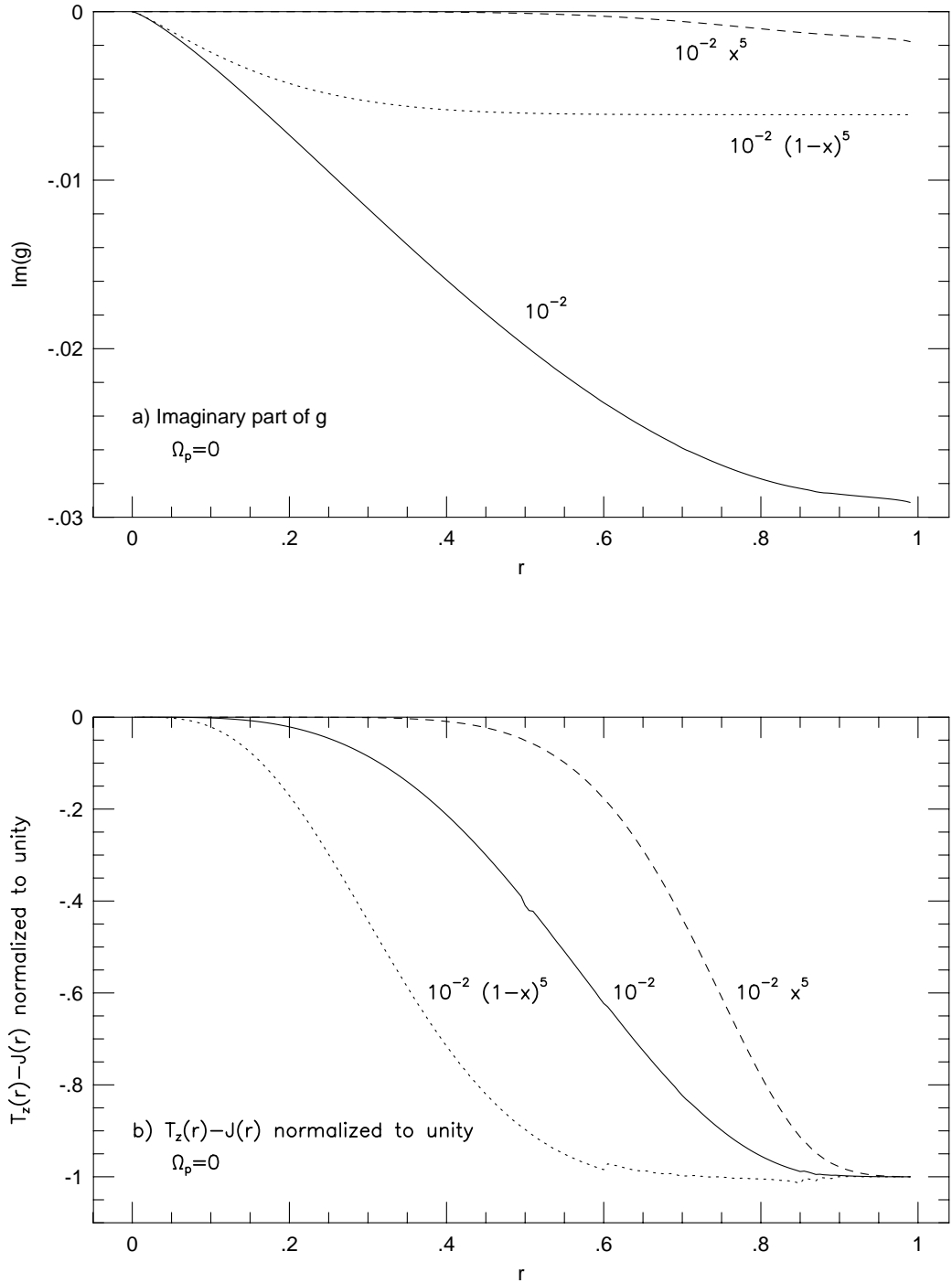}
\caption{ Imaginary part of $\tilde{g}$ ({\it panel a}) and
$T_z(r)-J(r)$ (angular momentum dissipated between $r_{in}$ and $r$)
normalized to unity ({\it panel b}) vs. $r$ for $\Omega_p=0$ and a
nonuniform $\alpha$. The different curves correspond to
$\alpha=10^{-2}$ ({\it solid line}), $10^{-2} \left( 1-x \right)^5$
({\it dotted line}) and $10^{-2} x^5$ ({\it dashed line}). The
parameters are $D=3$ and $(H/r)_{max}=0.1$.  }
\label{fig7} 
\end{figure}

Figure~\ref{fig7}.a shows the imaginary part of $\tilde{g}$ versus $r$
for $\Omega_p=0$ corresponding to these viscosity laws. The uniform
$\alpha_{\rm v}$ case is also plotted for comparison. The real part of
$\tilde{g}$ is not shown since it does not depend on $\alpha_{\rm v}$
for these small values of the viscosity (see discussion above). We see
that when the outer parts of the disk are nonviscous,
$Im(\tilde{g})$ in the inner parts is similar to the uniform
$\alpha_{\rm v}$ case, while it has an almost constant value in the
outer parts.  When the inner parts of the disk are nonviscous,
$Im(\tilde{g})$ is almost zero there, while it is significant in the
outer parts.  This behavior can be understood by remembering that
$Im[\tilde{g}(r)]$ is the integral from $r_{in}$ to $r$ of a function
proportional to $\alpha_{\rm v}$ (see equation~[\ref{zerog}]). Since
$Im(\tilde{g})$ varies less through the disk than in the uniform
$\alpha_{\rm v}$ case, $\tilde{v}'_r$ given by equation~(\ref{vr}) is
smaller.  This, together with a smaller $\alpha_{\rm v}$, leads to a
smaller net torque given by equation~(\ref{Tzapp}).
 
When $\Omega_p \ne 0$, there is not such a difference between the
uniform and nonuniform $\alpha_{\rm v}$ cases. The torque is smaller
in the nonuniform case, but $Im(\tilde{g})$ does not change
significantly from one case to the other. The reason is probably that
$Im(\tilde{g})$ does not vary monotonically with $\alpha_{\rm v},$ as
shown in Figure~\ref{fig1}.
 
Figure~\ref{fig7}.b shows $T_z(r)-J(r)$ versus $r$ for the same models
and $\Omega_p=0$.  We observe that, as expected, no angular momentum
is deposited at radii where $\alpha_{\rm v}$ is very small (small
radii for $\alpha_{\rm v} \propto x^5$ and large radii for
$\alpha_{\rm v} \propto (1-x)^5$).
 
\section{Discussion} 
\label{sec:discussion} 
 
In this paper, we have calculated the response of a viscous gaseous
disk to tidal perturbations with azimuthal mode number $m=1$ and odd
symmetry in $z$ with respect to the equatorial plane for both zero and
finite perturbing frequencies. We concentrated on these types of
perturbations because they arise for inclined disks and they lead to a
long--wavelength response. The effects of perturbations with even
symmetry, which occur in the coplanar case, may be superposed on the
effects of those with the odd symmetry.
 
Since the response of a viscous disk is not in phase with the 
perturbing potential, a tidal torque is exerted on the disk. When the 
perturber rotates outside the disk, this torque results in a decrease 
of the disk angular momentum. Part of the (negative) angular momentum 
of the perturbation is carried by tidal waves away from the location 
where the torque is exerted, and part is dissipated locally through 
viscous interaction of the waves with the background flow. 
 
We have shown that the tidal torque is comparable to the horizontal
viscous stress acting on the background flow when the perturbed
velocities in the disk are on the order of the sound speed $c_s$. If
these velocities remain subsonic, the tidal torque can exceed the
horizontal viscous stress only if the parameter $\alpha_{\rm v}$ which
couples to the vertical shear is larger than the parameter
$\alpha_{\rm h}$, which is coupled to the horizontal shear.  We note
that, so far, there is no indication about whether these two
parameters should be the same or not.  Nelson~\& Papaloizou (1998)
have found that, when the perturbed velocities exceed $c_s$, shocks
reduce the amplitude of the perturbation such that the disk moves back
to a state where these velocities are smaller than $c_s$.  When shocks
occur, the tidal torque exerted on the disk may become larger than the
horizontal viscous stress.
 
We have found that, in protostellar disks, bending waves are able to
propagate and transport a significant fraction of the negative angular
momentum they carry in the disk inner parts.  This is due to their
relatively large wavelength.  Therefore, tidal interactions in
noncoplanar systems may not be confined to the regions close to the
disk outer edge where the waves are excited.  For the disk models we
have set up, which extends down to the stellar surface, the value of
$\alpha_{\rm v}$ above which the waves get dissipated before reaching
the disk inner boundary varies between $5 \times 10^{-3}$ and
$10^{-2}$ for a disk aspect ratio between 0.05 and 0.1 (this critical
value of $\alpha_{\rm v}$ increases with the disk semi--thickness).
We note that if we had set up an annulus rather than a disk, the
surface mass density would drop at the inner boundary.  Then, for
small $\alpha_{\rm v}$, the waves would have a tendency to become
nonlinear and then to dissipate before reaching the inner boundary.
Thus, the behavior of bending waves could be similar in an annulus
with small $\alpha_{\rm v}$ and in a disk with larger $\alpha_{\rm
v}$.  In the limit of small viscosity, dissipation could also occur as
a result of parametric instabilities (Papaloizou~\& Terquem~1995;
C.F.~Gammie, J.J.~Goodman~\& G.I.~Ogilvie 1998, private
communication), the effect of which may be to lead to a larger
effective $\alpha_{\rm v}$.  However, since the waves we consider here
propagate in a nonuniform medium, it is not clear whether they would
be efficiently damped by these instabilities.
 
It the disk model allows at least partial reflection from the center,
the tidal interaction becomes resonant when the frequency of the tidal
waves is equal to that of some free normal global bending mode of the
disk.  For the equilibrium disk models we have considered, we have
found that a particularly strong resonance occurs when the separation
is about 8.5 times the disk radius.  The torque associated with the
finite frequency terms then increases by many orders of magnitude.
However the response may be limited by shocks and nonlinear effects.
The strength of the resonance is inversely proportional to
$\alpha_{\rm v}$.  In our calculations, the range of separation for
which the torque is significantly increased is rather large, which
means that the effects of tidal interaction, like disk truncation, may
occur even when the separation of the binary is large.  In addition to
this strong resonance we have found a few other resonances with
different strengths and widths.  From these calculations, it appears
that the probability for the interaction to be resonant may be
significant.  However, these results depend on the particular
equilibrium disk models we have set up.

We note that the disk is expected to be truncated such that the inner 
Lindblad (2:1) resonance with the companion, which could provide 
effective wave excitation, is more or less excluded from the disk.
 
Out of resonance, we find that the torque associated with the
zero--frequency perturbing term is much larger than that associated
with the finite frequency terms.  Of course, if the separation of the
system is large enough (at least 10 times the disk radius in our
models), the difference between finite frequency and secular terms
disappears.  When the secular response is dominant, the tidal torque
is proportional to $\alpha_{\rm v}$ in the limit $\alpha_{\rm v} \ll
1$. This has the consequence that the ratio of the tidal timescale
$t_d$ (time that would be required for the tidal effects we have
considered to remove the angular momentum content of the disk) to the
disk viscous timescale $t_{\nu , {\rm h}}$ is proportional to
$\alpha_{\rm h}/\alpha_{\rm v}$.

What observable effects would these tidal interactions produce in
protostellar disks ?  First, as we mentioned in the introduction, they
would lead to the precession of the jets that originate from these
disks.  The precession timescale that we can infer from the
observation of jets that are modeled as precessing is consistent with
this motion being induced by tidal effects in binary systems (Terquem
et al. 1998).  

Also, as we have already pointed out in \S~\ref{sec:validity}, the
secular perturbation produces a tilt the variation of which can be up
to $\alpha_{\rm v}$ along the line of nodes and $H^2/r^2$ along the
direction perpendicular to the line of nodes. Superposed on this
steady (in the precessing frame) tilt, there is a tilt produced by the
finite frequency perturbations, the variation of which across the disk
can be up to $H/r$.  Such asymmetries in the outer parts of
protostellar disks could be observed.  Because the variations along
the line of nodes and along the perpendicular to the line of nodes
depend on $\alpha_{\rm v}$ and $H/r$, respectively, observations of
warped protostellar disks have the potential to give important
information about the physics of these disks.  

Protostellar disks are believed to be rather thick, i.e. $H/r \sim
0.1$. Our calculations show that for such an aspect ratio, the
viscosity above which bending waves are damped before reaching the
central star is $\alpha_{\rm v} \sim 10^{-2}$. Observations seem to
indicate that $\alpha_{\rm h}$ in protostellar disks is on the order
of $10^{-3}$--$10^{-2}$.  If $\alpha_{\rm v}$ is smaller than or
comparable to $10^{-2}$, resonances may occur, as described above,
providing the disk inner edge allows some reflection of the waves.  In
that case, we could observe truncated disks with sharp edges even when
the binary separation is large.  If $\alpha_{\rm v}$ is larger than
but close to $10^{-2}$, bending waves are still able to propagate
throughout most of the disk.  In addition, given that protostellar
disks are rather thick, these waves propagate fast (they cross the
disk on a time comparable to the sound crossing time).  In that case,
it may be possible to observe some time--dependent phenomena with a
frequency equal to twice the orbital frequency.

In the case of protostellar disks, comparison between observations and
theory is now becoming possible. A noncoplanar binary system, HK~Tau,
has been observed for the first time very recently (see
\S~\ref{sec:intro1}), and subsequent work will be devoted to
interpreting these observations. In the meantime, we can comment
briefly on some observations related to X--ray binaries.
 
There is evidence from the light curve of X--ray binaries, such as
Hercules X--1 and SS~433, that their associated accretion disks may be
in a state of precession in the tidal field of the binary
companion. Katz~(1980a, 1980b) has indicated that the observed
precession periods of these two systems are consistent with the
precession being induced by the tidal field of the secondary. \\ In
the case of SS~433, it is interesting to note that an additional
``nodding'' motion with a period half the orbital period is observed
(see Margon~1984 and references therein). Katz~et al.~(1982) suggested
that this nodding motion is produced by the gravitational torque
exerted by the companion. In their model, the nodding is viscously
transmitted from the outside, where the torque is significant, to the
interior, where the jets are seen to respond to the motion. This
implies an extremely large viscosity in the disk. 

Here we point out that the observed period of this motion is
consistent with the nodding being produced by bending waves. As these
waves propagate with the sound speed and the disk is observed to be
very thick (Margon~1984), they could transmit the motion through the
disk on a timescale comparable to its dynamical timescale. For the
waves to reach the disk interior, $\alpha_{\rm v}$ would have to be
smaller than $H/r$. It is not clear that this condition is satisfied
here. The disk viscosity would then probably have to be smaller than
that predicted by the model of Katz~et al.~(1982).  

We finally observe that a rapid communication time through the disk
also makes problems with the radiation--driven warping model proposed
by Maloney~\& Begelman~(1997) to explain the precession of the disk in
SS~433. In this model, the communication between the different parts
of the disk, which is assumed to occur because of viscosity alone, is
on a timescale that is characteristic for mass to flow through the
disk. It is therefore very slow since the viscous timescale must be
long for the warping instability to occur (Pringle~1996).

\begin{acknowledgements} 
It is a pleasure to thank John Papaloizou for his advice,
encouragement, and many valuable suggestions and discussions. I
acknowledge Steve Balbus, Doug Lin, Jim Pringle and Michel Tagger for
very useful comments on an earlier draft of this paper, and John
Larwood for pointing out the ``nodding'' motion in SS~433.  I also
thank an anonymous referee whose comments helped to improve the
quality of this paper.  This work was supported by the Center for Star
Formation Studies at NASA/Ames Research Center and the University of
California at Berkeley and Santa Cruz.  I am grateful to the Isaac
Newton Institute at Cambridge University for support during the final
stages of this work.
\end{acknowledgements} 

\appendix

We describe here in more detail the resonance that occurs at $D \sim
8.5$, since it is particularly strong.  For $\Omega_p=2\omega$ and $-2
\omega$, resonance occurs at $D=8.25$ and $D=8.425$ respectively. We
observe that, for the values of $\alpha_{\rm v}$ considered here, the
range of $D$ for which the torque is significantly increased is rather
large. We see that the strength of the resonance is proportional to
$\alpha_{\rm v}^{-1}$. This is expected since the strength of the
resonance is inversely proportional to the damping factor, which
itself is proportional to $\alpha_{\rm v}$ when dissipation is due to
viscosity. We also note that the integral of the torque over the range
of frequencies in this resonance is independent of $\alpha_{\rm
v}$. This can be understood in the following way. For
$\Omega_p=2\omega$ (the same argument would apply for
$\Omega_p=-2\omega$), the torque at or close to resonance can be
written
 
\begin{equation} 
T_z = \frac{- {\cal A} \alpha_{\rm v}}{4 \left( \bar{\omega} -
\bar{\omega}_0 \right)^2 + \alpha_{\rm v}^2},
\label{T_app} 
\end{equation} 
 
\noindent where $\bar{\omega}=\omega/\Omega(R)$, $\omega_0$ is the
resonant frequency and ${\cal A}$ is an amplitude which does not
depend on $\alpha_{\rm v}$. We have calculated that ${\cal A} \simeq 8
\times 10^{-11}$. For the values of $\alpha_{\rm v}$ considered, we
have checked that the integral of the torque from
$\bar{\omega}_0-\Delta \bar{\omega}$ to $\bar{\omega}_0+\Delta
\bar{\omega}$ does not depend very much on whether $2 \Delta
\bar{\omega}$ is taken to be the range of frequencies in the resonance
or for which expression~(\ref{T_app}) is valid (the latter being
smaller). We then approximate the integral $I$ of the torque over the
resonance by using~(\ref{T_app}), so that
 
\be
I = - {\cal A} \; Arctan \frac{2 \Delta \bar{\omega}}{\alpha_{\rm v}}
.
\ee
 
\noindent For $\alpha_{\rm v}=10^{-3}$ or $10^{-4}$, the range of
frequencies $2 \Delta \bar{\omega}$ for which~(\ref{T_app}) is valid
is large compared with $\alpha_{\rm v}$, so that $I \simeq - {\cal A}
\pi/2$, independent of $\alpha_{\rm v}$. For $\alpha_{\rm v}=10^{-2}$,
although $2 \Delta \bar{\omega}$ is only a few times $\alpha_{\rm v}$,
this relation is still valid within a factor less than 2 (due to the
rapid convergence of the function $Arctan$).

\newpage 
 


\begin{table}[ht] 
\footnotesize
\caption[]{Torque and Corresponding Tidal Evolution Timescale.\\}
\begin{tabular}{ccccccccccccccccccccccccccccccccc} 
\tableline 
\tableline 
Label & $D$ & $\alpha_{\rm v}$ & $\left( \frac{H}{r} \right)_{max}$ &\
& & $T_z$ & &\ & $\omega_p$ & $t_d/t_{\nu , {\rm h}}$ \\
\cline{6-8} 
& & & & & $2\omega$ & $-2\omega$ & 0 & & ($10^{-2}$) & \\ 
\tableline 
1a & 3 & $10^{-4}$ & 0.1 & & $-4 (3) \times 10^{-12}$
\tablenotemark{a} & $ -2 (0.8) \times 10^{-12} $ & $-1 (1)
\times 10^{-9} $ \tablenotemark{b} & & -1.1 & 5.9 \\
1b & -- & $10^{-3}$ & -- & & $-4 (3) \times 10^{-11} $
\tablenotemark{a} & $ -2 (0.7) \times 10^{-11} $ & $-1 (1) \times
10^{-8} $ \tablenotemark{b} & & -- & -- \\
1c & -- & $ 10^{-2}$ & -- & & $-3 (2) \times 10^{-10} $
\tablenotemark{a} & $ -1 (0.5) \times 10^{-10} $ & $-1 (1) \times
10^{-7}$ \tablenotemark{c} & & -- & -- \\
1d & -- & $0.1$ & -- & & $-9 (8) \times 10^{-10} $ \tablenotemark{a} &
$\ \ \ -4 (2) \times 10^{-10} $ & $-1 (1) \times 10^{-6} $
\tablenotemark{c} & & -- & -- \\
& & & & & & & & & & \\ 
2a & 6 & $10^{-3}$ & 0.05 & & $-2 (2) \times 10^{-11} \ $ & $ \ \ \ -3
(2) \times 10^{-12}$ & $\ -8 (6) \times 10^{-10} \ $ & & -0.14 & 22.9 \\
2b & -- & $5 \times 10^{-3}$ & -- & & $-1 (0.7) \times 10^{-10} \ \ \
$ & $-1 (0.8) \times 10^{-11}$ & $ -4 (3) \times 10^{-9} \ $ & & --
& 23.1 \\
2c & -- & $10^{-2}$ & -- & & $-2 (1) \times 10^{-10} \ $ & $\ \ \ 
-2 (1) \times 10^{-11}$ & $ -8 (6) \times 10^{-9} \ $ & & -- & 23.2
\\
2d & -- & $5 \times 10^{-2}$ & -- & & $ -3 (2) \times 10^{-10} \ $ &
$\ \ \ -3 (2) \times 10^{-11}$ & $-4 (3) \times 10^{-8} \ $ & & -- &
23.5 \\
2e & -- & $0.1$ & -- & & $-3 (2) \times 10^{-10} \ $ & $\ \ \ -3 (2)
\times 10^{-11}$ & $-8 (7) \times 10^{-8} \ $ & & -- & 23.6 \\
& & & & & & & & & & \\ 
3a & 4 & $10^{-3}$ & 0.1 & & $-5 (4) \times 10^{-11} \ $ & $\ \ \ -2
(1) \times 10^{-11} $ & $\ -2 (2) \times 10^{-9} \ $ \tablenotemark{a}
& & -0.47 & 32.3 \\
3b & 5 & -- & -- & & $-4 (3) \times 10^{-11} \ $ & $\ \ \ -8 (5)
\times 10^{-12} $ & $\ -6 (5) \times 10^{-10} $ \tablenotemark{a} & &
-0.24 & 118.3 \\
3c & 6 & -- & -- & & $-7 (5) \times 10^{-11} \ $ & $-1 (0.8) \times
10^{-11} $ & $\ -2 (2) \times 10^{-10} $ \tablenotemark{a} & & -0.14 &
274.9 \\
3d & 7 & -- & -- & & $-2 (1) \times 10^{-10} \ $ & $-1 (0.8) \times
10^{-11} $ & $\ -8 (6) \times 10^{-11} $ \tablenotemark{a} & & -0.087
& 287.8 \\
3e & 8 & -- & -- & & $-4 (3) \times 10^{-9} \ \ $ & $\ \ \ -9 (6)
\times 10^{-11} $ & $\ -4 (3) \times 10^{-11} $ \tablenotemark{a} & &
-0.058 & 21.9 \\
3f & 8.25 & -- & -- & & $-7 (6) \times 10^{-8} \ \ $ & $\ \ \ -4 (3)
\times 10^{-10} $ & $\ -3 (2) \times 10^{-11} $ \tablenotemark{a} & &
-0.053 & 1.1 \\
3g & 8.425 & -- & -- & & $-7 (5) \times 10^{-9} \ \ $ & $\ \ -4 (3)
\times 10^{-9} $ & $\ -3 (2) \times 10^{-11} $ \tablenotemark{a} & &
-0.050 & 7.8 \\
3h & 9 & -- & -- & & $-3 (3) \times 10^{-10} \ $ & $\ \ \ -4 (3)
\times 10^{-11} $ & $\ -2 (1) \times 10^{-11} $ \tablenotemark{a} & &
-0.041 & 201.0 \\
3i & 10 & -- & -- & & $-5 (4) \times 10^{-11} \ $ & $\ \ \ -4 (3)
\times 10^{-12} $ & $-1 (0.7) \times 10^{-11} $ \tablenotemark{a} & & 
-0.030 & 1235.1 \\
& & & & & & & & & & \\ 
4a & 4 & $10^{-2}$ & -- & & $-5 (4) \times 10^{-10} \ $ & $ -1 (0.8)
\times 10^{-10}$ & $-2 (2) \times 10^{-8} \ $ & & -0.47 & 32.4 \\
4b & 5 & -- & -- & & $-4 (3) \times 10^{-10} \ $ & $\ \ \ -7 (5)
\times 10^{-11}$ & $-6 (5) \times 10^{-9} \ $ & & -0.24 & 118.7 \\
4c & 6 & -- & -- & & $-6 (5) \times 10^{-10} \ $ & $\ \ \ -9 (6)
\times 10^{-11}$ & $-2 (2) \times 10^{-9} \ $ & & -0.14 & 280.4 \\
4d & 7 & -- & -- & & $-2 (1) \times 10^{-9} \ \ $ & $-1 (0.7) \times
10^{-10}$ & $-8 (6) \times 10^{-10} \ $ & & -0.087 & 315.8 \\
4e & 8 & -- & -- & & $-7 (5) \times 10^{-9} \ \ $ & $\ \ \ -3 (2)
\times 10^{-10}$ & $-4 (3) \times 10^{-10} $ & & -0.058 & 110.8 \\
4f & 8.25 & -- & -- & & $-7 (6) \times 10^{-9} \ \ $ & $\ \ \ -4 (3)
\times 10^{-10}$ & $-3 (2) \times 10^{-10} $ & & -0.053 & 101.2 \\
4g & 8.425 & -- & -- & & $-6 (5) \times 10^{-9} \ \ $ & $\ \ \ -4 (3)
\times 10^{-10}$ & $-3 (2) \times 10^{-10} $ & & -0.050 & 117.1 \\
4h & 9 & -- & -- & & $-2 (2) \times 10^{-9} \ \ $ & $\ \ \ -2 (1)
\times 10^{-10}$ & $-2 (1) \times 10^{-10} $ & & -0.041 & 323.9 \\
4i & 10 & -- & -- & & $ -5 (3) \times 10^{-10} \ $ & $\ \ \ -3 (2)
\times 10^{-11}$ & $-1 (0.7) \times 10^{-10} \ \ $ & & -0.030 & 1368.7
\\
\tableline 
\end{tabular} 
\tablecomments{Listed are the separation of the system $D$, the
viscous parameter $\alpha_{\rm v}$, the maximum value of $H/r$, the
torque $T_z$ associated with each of the perturbed frequencies $2
\omega$, $-2 \omega$ and 0, the disk precession frequency $\omega_p$
and the ratio of the tidal timescale $t_d$ to the viscous timescale
$t_{\nu , {\rm h}}$.}
\tablenotetext{a}{Cases where,at the outer edge of the disk, $g$
varies on a scale smaller than $H$. }
\tablenotetext{b}{Cases where,at the outer edge of the disk, the
radial spherical velocity is larger than the sound speed.}
\tablenotetext{c}{Cases where,at the outer edge of the disk, $g$
varies on a scale smaller than $H$ {\it and} the radial spherical
velocity is larger than the sound speed.}
\label{table1} 
\end{table} 

\newpage 

\end{document}